\documentclass{vldb}
\usepackage{graphicx}
\usepackage{balance}  % for  \balance command ON LAST PAGE  (only there!)
\usepackage{hyperref}
\usepackage{listings}
\usepackage{courier}
\usepackage[T1]{fontenc}
\usepackage{cleveref}
\usepackage{amsmath}
\usepackage{amssymb}
\usepackage{latexsym}
\usepackage{times}
\usepackage{bm}
\usepackage{color}
\usepackage{subscript}
\usepackage{multirow}
\usepackage{kbordermatrix}
\usepackage{booktabs} % For formal tables
\usepackage{algorithm}
\usepackage{mathtools}
\usepackage{commath}
\usepackage[noend]{algpseudocode}
\usepackage{array}
\usepackage[page]{appendix}
\usepackage{url}

\usepackage[labelfont=bf]{caption}

\makeatletter
\def\BState{\State\hskip-\ALG@thistlm}
\makeatother

\newcommand\learningParameter{\Theta}
\newcommand\tuple{\textbf{t}}
\newcommand\lossFunction{L}

\lstnewenvironment{SQL}
  {\lstset{
        aboveskip=5pt,
        belowskip=5pt,
        escapechar=!,
        mathescape=true,
        language=SQL,
        basicstyle=\linespread{0.94}\ttfamily\small,
        morekeywords={BULK, COLLECT, FOR, MATRIX, VECTOR, EACH, INTEGER, LONG, DOUBLE, WITH, PARTITION, TEST, WHETHER, PROBABILITY, VECTORIZE, ROWMATRIX, inner\_product, matrix\_vector\_multiply, MATRIX\_MULTIPLY, outer\_product, matrix\_inverse, trans\_matrix,
        label\_scalar, label\_vector, get\_scalar, diag, rowMins, rowIndexMax, map, transpose, multiply, asInstanceOf, DenseMatrix, reduce,
        toArray, zipped, toIndexedRowMatrix, if, Tuple2, override, def, new, gemm, build, filter, matmul, crossentropyderiv, reluderiv, t, relu, softmax, reducebyrow},
        deletekeywords={VALUE, PRIOR},
        showstringspaces=true}
        \vspace{0pt}%
        \noindent\minipage{0.47\textwidth}}
  {\endminipage\vspace{0pt}}

\vldbTitle{Declarative Recursive Computation on an RDBMS}
\vldbAuthors{Dimitrije Jankov, Shangyu Luo, Binhang Yuan, Zhuhua Cai, Jia Zou, Chris Jermaine, and Zekai J. Gao}
\vldbDOI{https://doi.org/10.14778/3317315.3317323}
\vldbVolume{12}
\vldbNumber{7}
\vldbYear{2019}

\begin{document}

% ****************** TITLE ****************************************

\title{Declarative Recursive Computation on an RDBMS}

% possible, but not really needed or used for PVLDB:
\subtitle{or, Why You Should Use a Database For Distributed Machine Learning}
%\titlenote{A full version of this paper is available as\textit{Author's Guide to Preparing ACM SIG Proceedings Using \LaTeX$2_\epsilon$\ and BibTeX} at \texttt{www.acm.org/eaddress.htm}}}

% ****************** AUTHORS **************************************


\author{
    Dimitrije Jankov{\fontsize{12}{14}$^\dagger$}, 
    Shangyu Luo{\fontsize{12}{14}$^\dagger$},
    Binhang Yuan{\fontsize{12}{14}$^\dagger$}, 
  \\Zhuhua Cai{\fontsize{12}{14}*}, 
    Jia Zou{\fontsize{12}{14}$^\dagger$}, 
    Chris Jermaine{\fontsize{12}{14}$^\dagger$}, 
    Zekai J. Gao{\fontsize{12}{14}$^\dagger$}
    \\
    {\affaddr{Rice University {\fontsize{12}{14}$^\dagger$*}}}
    \\ 
    \text{\{dj16, sl45, by8, jiazou, cmj4, jacobgao\}}@rice.edu $^\dagger$\\
    \text{caizhua@gmail.com *}
}
% There's nothing stopping you putting the seventh, eighth, etc.
% author on the opening page (as the 'third row') but we ask,
% for aesthetic reasons that you place these 'additional authors'
% in the \additional authors block, viz.
%\additionalauthors{}
% \date{30 July 1999}
% Just remember to make sure that the TOTAL number of authors
% is the number that will appear on the first page PLUS the
% number that will appear in the \additionalauthors section.

\maketitle

\begin{abstract}
A number of popular systems, most notably Google's TensorFlow,
have been implemented from the
ground up to support machine learning tasks.
{\color{black} We
consider how to make a very small set of changes to a
modern relational database management system (RDBMS) to make it suitable for distributed learning computations.  Changes include adding
better support for recursion, and optimization and execution of very large compute plans.}
We also show that there are key advantages to using an RDBMS as a machine learning platform.  In particular, learning based on a database management system allows for trivial scaling to large data sets and especially large models, where different computational units operate on different parts of a model that may be too large to fit into RAM.
\end{abstract}

\section{introduction}

Modern machine learning (ML) platforms such as TensorFlow \cite{tensorflow} have primarily been designed to support \emph{data parallelism}, where
a set of almost-identical computations (such as the computation of a gradient) are executed in parallel over a set of computational units.
The only difference among the computations is that each operates over different training data (known as ``batches'').  After each computation has
finished, the local result is either loaded to a parameter server (in the case of asynchronous data parallelism \cite{recht2011hogwild}) or 
the local results are globally aggregated and used to update the model (in the case of synchronous data parallelism \cite{hillis1986}).

Unfortunately, data parallelism has its limits.  For example, data parallelism implicitly
assumes that the model being learned (as well as
intermediate data produced when a batch is used to update the model) can fit in the RAM of a computational 
unit (which may be a server machine or a GPU).  This is not always a reasonable assumption, however.
For example, a
state-of-the-art NVIDIA Tesla V100 Tensor Core GPU (a \$10,000 data center GPU) has 32GB of RAM. 32GB of RAM cannot
store the matrix required for a fully-connected layer to
encode a vector containing entries from 200,000 categories into a vector of 50,000 neurons.
Depending upon the application, 50,000 neurons may not be a lot \cite{shazeerMMDLHD17}.

Handling such a model requires \emph{model parallelism}---where the statistical 
model being learned is not simply replicated at different 
computational units, but is instead partitioned and operated over in parallel, {\color{black} and is executed by a series of bulk-synchronous operations. As
discussed in the related work section, existing} 
systems for distributed ML offer limited support for 
model parallelism.  

\vspace{5 pt}
\noindent
\textbf{Re-purposing relational technology for ML.}
We argue that
model parallelism can be implemented using relational technology.
Different parts of the model can be stored in a set of tables,
and the computations on the partial model can often be expressed through a few SQL queries.
In fact, to a programmer, the model-parallel SQL implementation of a learning algorithm looks no different than the
data parallel implementation.  
Relational database management systems (RDBMSs) 
provide a {\color{black}declarative programming interface},
which means that the programmer (or automated algorithm generator, if a ML algorithm is automatically generated via automatic
differentiation) only needs to specify
what he/she/it wants, but does not need to write out how to compute it.
The computations will be automatically generated by the system, 
and then be optimized and executed to match the data size, layout, and the compute hardware.
The code is the same whether the computation is run on a
local machine or in a distributed environment. 
In contrast, systems such as TensorFlow provide relatively weak forms of {\color{black}declarative-ness}, as
each logical operation in a compute graph (such as a matrix multiply) must be specified and executed on some physical compute unit, like a GPU.

Another benefit of using relational technology is that
distributed computations in RDBMSs have been studied for more than thirty years,
and are fast and robust. 
The query optimizer, shipped with an RDBMS, 
is highly effective for optimizing distributed computations \cite{chaudhuri1998overview}.
It is not an accident that competing distributed compute platforms
such as Spark \cite{zaharia2010spark} (which now promotes the use of relational-style DataFrames \cite{Armbrust:2015} and DataSets \cite{datasets} interfaces) 
are beginning to look more like a parallel RDBMSs.

\vspace{5 pt}
\noindent
\textbf{Challenges of adapting RDBMS technology for ML.} 
However, there are a couple of reasons that a
modern RDBMS cannot be used out-of-the-box as a platform for most large-scale ML algorithms.  
Crucially, such systems lack sufficient
support for recursion. In deep learning it is necessary to ``loop'' through the layers
of a deep neural network, and then ``loop'' backwards through the network
to propagate errors.
Such ``looping'' could be expressed declaratively via recursive dependencies among tables, but RDBMS 
support for recursion is typically limited (if it exists at all)
to computing fixed-points over sets such as transitive closures \cite{Aho1979TheUO}.
Not only that, but there is the problem that
the query plan for a typical deep-learning computation may run to tens of thousands of operators, which no 
existing RDBMS optimizer is going to be able to handle.

\vspace{5 pt}
\noindent
\textbf{Our Contributions.}
Specific contributions are:

\vspace{-3 pt}
\begin{itemize}
\item We introduce \emph{multi-dimensional}, \emph{array-like} indices to databa-se tables. 
When a set of tables share the similar computation pattern, 
they can be compacted and replaced by a table with multiple versions (indicated by its indices).

\vspace{-5 pt}
\item We modify the query optimizer of a database to render it capable of handling very large query graphs. 
A query graph is partitioned into a set of runnable \emph{frames}, and the cost of operators and pipelining are considered. 
We formalize the graph-cutting problem as 
an instance of the \emph{generalized quadratic assignment problem} \cite{lee:2004}.

\vspace{-5 pt}
\item We implement our ideas on
top of SimSQL \cite{cai2013simulation}, which is a prototype distributed RDBMS that is specifically designed to handle large-scale statistical
computation.

\vspace{-5 pt}
\item We test our implementations on two distributed deep learning problems 
(a feed-forward neural network and an implementation of Word2Vec \cite{word2vec:NIPS2013, Mikolov2013EfficientEO})
{\color{black} as well as distributed latent Dirichlet allocation (LDA).
We show that declarative SimSQL codes  
scale to huge model sizes, past the model sizes that TensorFlow can support,
and that SimSQL can outperform TensorFlow on some models.}
\end{itemize}

\section{Parallelism in ML}

Because one of the key benefits of ML on an RDBMS is automated parallelism, we begin with a brief
review of parallelism in ML.

In the general case, when solving a ML problem, we are given a data set
$\textbf{T}$ with elements $\textbf{t}_j$.  The goal is to learn a
$d$-dim\-en\-sion\-al vector ($d\ge 1$) of model parameters
$\learningParameter=(\learningParameter^{(1)}$,
$\learningParameter^{(2)},\dots,\learningParameter^{(d)})$ that minimize a
loss function of the form $\sum_j \lossFunction(\tuple_j |
\learningParameter)$. To this end, learning algorithms such as gradient descent perform a simple
update
repeatedly until convergence:
\[
\learningParameter_{i + 1} \gets \learningParameter_i - 
F (\learningParameter_i, \textbf{T})
%\nabla\lossFunction(\tuple_j | \learningParameter_i).
\]
\noindent Here, $F$ is the \emph{update function}. 
Each update marks the end of a processing
\emph{epoch}.
Many learning algorithms are \emph{decomposable}.  That is, 
if $\textbf{T}$ has elements $\textbf{t}_j$, the algorithm can be written as:
\[
\learningParameter_{i + 1} \gets \learningParameter_i - 
\sum_j F (\learningParameter_i, \textbf{t}_j)
%\nabla\lossFunction(\tuple_j | \learningParameter_i).
\]
For example, consider gradient descent, the quintessential learning algorithm.  It is decomposable because
$F (\learningParameter_i, \textbf{T}) = 
\sum_j \nabla\lossFunction(\tuple_j |
\learningParameter_i)$.

If it is possible to store $\learningParameter_i$ in the RAM of each machine,
decomposable learning algorithms can be made \emph{data parallel}.  One can broadcast
$\learningParameter_i$ to each site, and then compute $F (\learningParameter_i, \textbf{t}_j)$ for
data $\textbf{t}_j$ stored locally.  All of these values are then aggregated using standard, distributed
aggregation techniques.

However, data parallelism of this form is often ineffective. 
Let $\textbf{T}_i$ be a small sample of $\textbf{T}$ selected to compute the $i$th gradient update.
For decomposable algorithms, $F (\learningParameter_i, \textbf{T}) \approx \frac{|\textbf{T}|}{|\textbf{T}_i|}
F (\learningParameter_i, \textbf{T}_i)$, therefore in practice only a small subsample of  
the data are used (for example, in the case of gradient descent, \emph{mini-batch gradient descent} \cite{ruder2016overview} is typically used).   
Adding more machines can either distribute this sample so that each machine
gets a tiny amount of data (which is typically not helpful because for very small data sizes, the
fixed costs associated with broadcasting $\learningParameter_i$ dominate) or else use a larger sample.  This is
also not helpful because the estimate to $F (\learningParameter_i, \textbf{T})$ with a relatively small sample
is already accurate enough.  The largest batches advocated in the literature consist of around 10,000 samples \cite{goyal2017accurate}.

One idea to overcome this is to use \emph{asynchronous data parallelism} \cite{recht2011hogwild}, where recursion of the form
$\learningParameter_{i + 1} \gets \learningParameter_i -
F (\learningParameter_i, \textbf{T})$ is no longer used.  Rather, each site $j$ is given a small sample
$\textbf{T}_{j}$ of $\textbf{T}$; it requests the value $\learningParameter_{cur}$,
computes $\learningParameter_{new} \gets \learningParameter_{cur} -
F (\learningParameter_{cur}, \textbf{T}_j)$ and registers $\learningParameter_{new}$ at a parameter server.
All requests for $\learningParameter_{cur}$
happen to obtain whatever the last value written was, leading to stochastic behavior.
The problem is that data parallelism of this form can be ineffective for large computations as 
most of the computation is done using
stale data \cite{chen2016revisiting}.

An alternative is \emph{model parallelism}.  In model parallelism, the idea is to stage
$F (\learningParameter_i, \textbf{T})$ (or $F (\learningParameter_i, \textbf{T}_i)$)
as a distributed computation without assuming that each 
site has
access to all of $\learningParameter_i$ (or $\textbf{T}_i$).
There are many forms of model parallelism, but in the general case, model parallelism is ``distributed computing complete''.
That is, it is as hard as ``solving'' distributed computing.

The distributed key-value stores (known as \emph{parameter servers}) favored by most existing Big Data ML systems (such
as TensorFlow and Petuum \cite{Xing:2015}) make it difficult to build model parallel computations, even ``by hand''.
In practice, an operation such as a distributed matrix multiply on TensorFlow---a key building block for model parallel
computations---requires
a series of machine- and data-set- specific computational graphs to be constructed and executed, where communication is facilitated
by explicitly storing and retrieving intermediate results from the key-value store.
This is far enough outside of norm of how TensorFlow is designed to be used given that (at least at the time of this writing) no widely-used
codes for
distributed matrix multiplication on top of the platform exist.

\section{Deep Learning on an RDBMS}

\subsection{A Simple Deep Learner}

A deep neural network is a differentiable, non-linear function, typically conceptualized as a directed graph.  Each 
node in the graph (often called a ``neuron'') computes a continuous activation function over its inputs
(sigmoid, ReLU, etc.).
  
 \vspace{-5pt}
 
\begin{figure}[h]
\begin{center}
\includegraphics[width=7cm]{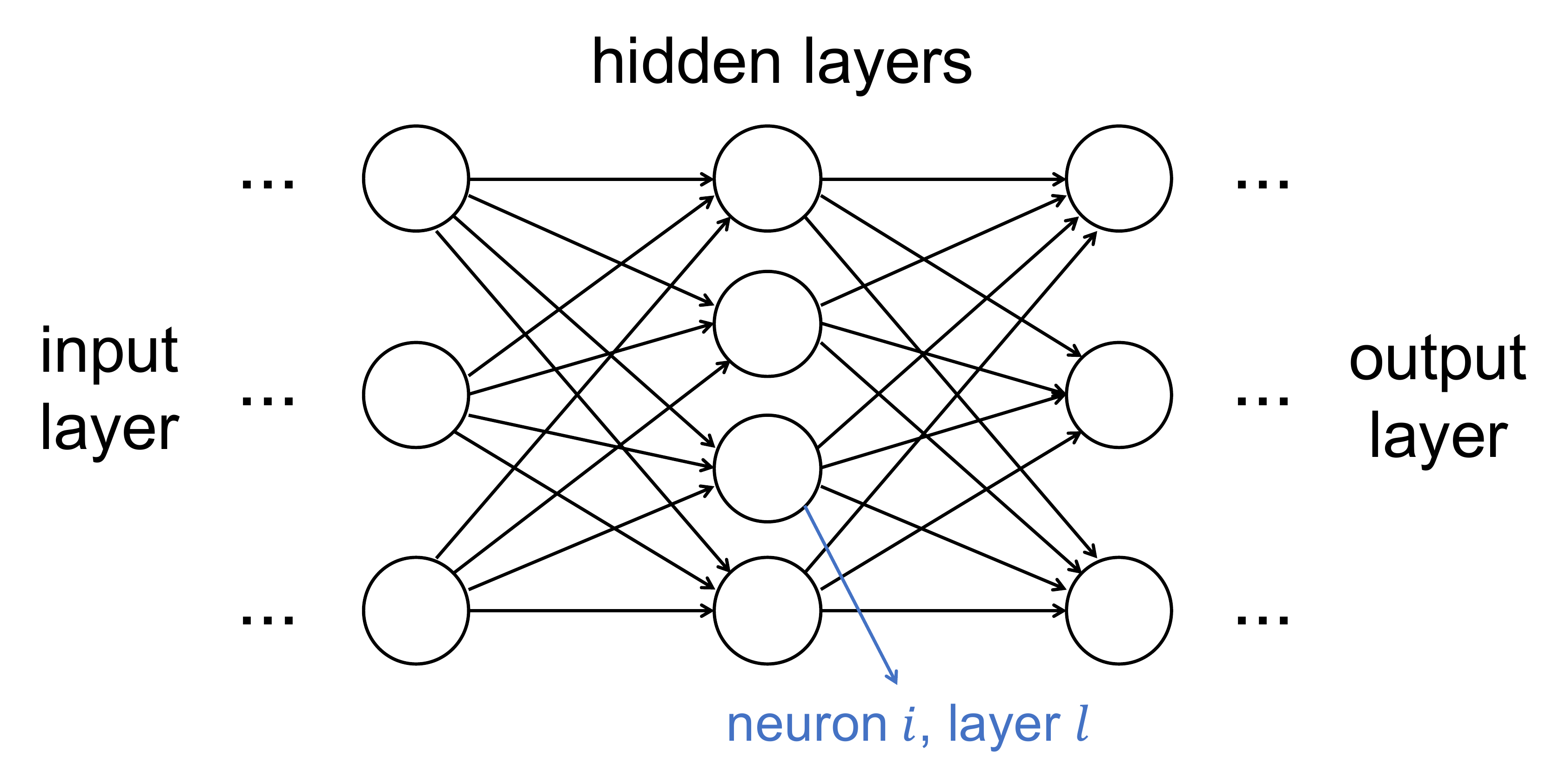}
\end{center}
\vspace{-10 pt}
 \caption{Structure of a feed-forward neural network.}
 \label{fig:ffnn}
\end{figure}

\vspace{-5pt}

One of the simplest and most commonly used artificial 
neural networks is a so-called \emph{feed-forward neural network} \cite{hornik1989multilayer}.
Neurons are organized into layers.  Neurons in one layer
are connected only to neurons in the next layer, hence the name ``feed-forward".  
{\color{black} Consider the feed-forward network in Figure~\ref{fig:ffnn}.}
To compute a function over an input (such as a text document or an image), the input vector is fed into the first layer, and the output
from that layer is fed through one or more hidden layers, until the output layer is reached.  
If the output of layer $l-1$ (or ``activation'') is represented as a vector $\textbf{a}_{l-1}$, then the output of layer
$l$ is computed as
$\textbf{a}_l=\sigma\left(\textbf{a}_{l-1}\textbf{W}_l+\textbf{b}_l\right)$
%Here, $\textbf{b}_l$ is the bias vector associated with layer $l$, $\textbf{W}_l$ is the weight matrix associated with the layer,
Here, $\textbf{b}_l$ and $\textbf{W}_l$ are the the bias vector and the weight matrix associated with the layer $l$, respectively, 
and $\sigma(\cdot)$ is the activation function.

\vspace{5 pt}
\noindent \textbf{Learning.}
\emph{Learning} is the process of customizing the weights for a particular data set and task.  Since learning is by far the most
computationally intensive part of using a deep network, and because the various data structures (such as the
$\textbf{W}_l$ matrix) can be huge, 
this is the part we would typically like to distribute across machines.

Two-pass mini-batch gradient descent is the
most common learning method used with such networks.
Each iteration takes as input the current set of weight matrices $\{\textbf{W}^{(i)}_1, \textbf{W}^{(i)}_2, ...\}$ and
bias vectors $\{\textbf{b}^{(i)}_1, \textbf{b}^{(i)}_2, ...\}$
and then outputs the next set of  weight matrices $\{\textbf{W}^{(i+1)}_1, \textbf{W}^{(i+1)}_2, ...\}$ and
bias vectors $\{\textbf{b}^{(i+1)}_1, \textbf{b}^{(i+1)}_2, ...\}$.
This process is repeated until convergence.

In one iteration of the gradient descent, each batch of inputs goes through two passes: the forward pass and the backward pass.

\vspace{5 pt}
\noindent \textbf{The forward pass.} In the forward pass, at iteration $i$,
a small subset of the training data are randomly selected and stored in
the matrix $\textbf{X}^{(i)}$.  The activation matrix for each of these data points, $\textbf{A}_{1}$, is computed as 
$\textbf{A}^{(i)}_1=\sigma\left(\textbf{X}^{(i)}\textbf{W}^{(i)}_1+\textbf{B}^{(i)}_1\right)$ (here, let the 
bias matrix $\textbf{B}^{(i)}_1$ be the matrix formed by replicating the bias vector $\textbf{b}^{(i)}_1$ $n$ times, where $n$ is the size
of the mini-batch).
Then, this activation is pushed through the network by repeatedly performing the computation
$\textbf{A}^{(i)}_{l}=\sigma\left(\textbf{A}^{(i)}_{l-1}\textbf{W}^{(i)}_l+\textbf{B}^{(i)}_l\right)$.

\vspace{5 pt}
\noindent \textbf{The backward pass.} At the end of the forward pass, a loss (or error function) comparing the predicted
set of values to the actual labels from the training data are computed.  
To update the weights and biases using gradient descent,
the errors are fed back through the network, using the chain rule.
Specifically, the errors back-propagated from hidden layer $l+1$  to layer $l$ in the $i$-th backward pass is computed as
$$\textbf{E}_l^{(i)}=\left(\textbf{E}_{l+1}^{(i)}\left(\textbf{W}_{l+1}^{(i)}\right)^T\right)
\odot\sigma'\left(\textbf{A}_l^{(i)}\right),$$ 

\noindent
where $\sigma'(\cdot)$ is the derivative of the activation function.
After we have obtained the errors (that serve as the gradients) for each layer, we update the weights and biases:
$$\textbf{W}_l^{(i)}=\textbf{W}_l^{(i-1)}-\alpha\cdot \textbf{A}_{l-1}^{(i-1)}\textbf{E}_l^{(i-1)},$$ 
$$\textbf{b}_l^{(i)}=\textbf{b}_l^{(i-1)}-\alpha\cdot \sum\limits_n\textbf{e}_l^{(i-1)},$$ 

\noindent
where $\alpha$ is the learning rate, and $\textbf{e}_l$ is the row vector of $\textbf{E}_l$.

\subsection{A Mixed Imperative/Declarative Approach}
{\color{black} Perhaps surprisingly, a model parallel version of the algorithm is possible on top of an RDBMS.
We assume that an RDBMS has been lightly augmented to handle \texttt{matrix} and \texttt{vector} data types 
as described in \cite{luo2017scalable}, and
assume that the various matrices and vectors have been ``chunked''. 
The following database table
stores the chunk of $\textbf{W}_{\texttt{LAYER}}^{(\texttt{ITER})}$ at the given row and column:

\vspace{5 pt}
\noindent 
\texttt{W (ITER, LAYER, ROW, COL, MAT)}

\vspace{5 pt}
\noindent \texttt{MAT} is of type \texttt{matrix (1000, 1000)}
and stores one ``chunk'' of $\textbf{W}_{\texttt{LAYER}}^{(\texttt{ITER})}$.  
A $10^5 \times 10^5$ matrix chunked in this way would have
$10^4$ entries in the table \texttt{W}, with one sub-matrix for each of the $100 = 10^5 / 10^3$ possible \texttt{ROW} values combined
with each of the $100 = 10^5 / 10^3$ possible \texttt{COL} values.

Also, the activations $\textbf{A}^{(\texttt{ITER})}_{\texttt{LAYER}}$ are chunked and stored as matrices having 1000 columns 
in the following table:

\vspace{5 pt}
\noindent
\texttt{A (ITER, LAYER, COL, ACT)}

\vspace{5 pt}
\noindent
A final table \texttt{AEW} stores the values needed to compute $\textbf{W}_{\texttt{LAYER}}^{(\texttt{ITER+1})}$:
$\textbf{A}^{(\texttt{ITER})}_{\texttt{LAYER-1}}$ (as \texttt{ACT}), $\textbf{E}^{(\texttt{ITER})}_{\texttt{LAYER}}$ (as \texttt{ERR}), and 
$\textbf{W}^{(\texttt{ITER})}_{\texttt{LAYER}}$ (as \texttt{MAT}):

\vspace{5 pt}
\noindent
\texttt{AEW (LAYER, ROW, COL, ACT, ERR, MAT)}

\vspace{5 pt}
\noindent
\texttt{ROW} and \texttt{COL} again identify a particular matrix chunk. 
Given this, a fully model parallel implementation of
the backward pass can be implemented using the SQL code in Figure~\ref{fig:backprop}.
\texttt{crossentropy\-deriv()} and \texttt{reluderiv()} are user-defined functions implementing the derivatives of cross-entropy and ReLU activation, respectively.
The %entire (All, I remove this word simply to shorten this paragraph so that we can remove an orphan line in the next paragraph. -Jia)
model parallel
backward-pass code is around twenty lines long and could be generated by an auto-differentiation tool.}

\begin{figure}[t]
\begin{SQL}
--First, issue a query that computes the errors
--being backpropagated from the top layer in
--the network.
SELECT 9, W.ROW, W.COL, A.ACT, E.ERR, W.MAT
BULK COLLECT INTO AEW
FROM A, W, 
  --Note: we are using cross-entropy loss
 (SELECT A.COL, 
         crossentropyderiv(A.ACT, DO.VAL) AS ERR
  FROM A, DATA_OUTPUT AS DO
  WHERE A.LAYER=9) AS E
WHERE A.COL=W.ROW AND W.COL=E.COL 
  AND A.LAYER=8 AND W.LAYER=9
  AND A.ITER=i AND W.ITER=i;
 
--Now, loop back through the layers in the network
for l = 9, ..., 2:
  --Use the errors to compute the new weights 
  --connecting layer l to layer l + 1; add to 
  --result for learning iteration i + 1 
  SELECT i+1, l, ROW, COL, 
         MAT - matmul(t(ACT), ERR) * 0.00000001
  BULK COLLECT INTO W
  FROM AEW WHERE LAYER=l;
  
  --Issue a new query that uses the errors from the
  --previous layer to compute the errors in this 
  --layer. reluderiv takes the derivative of the 
  --activation.
  SELECT l-1, W.ROW, W.COL, A.ACT, E.ERR, W.MAT
  BULK COLLECT INTO AEW FROM A, W,
   (SELECT ROW AS COL, SUM(matmul(ERR, t(MAT))
        * reluderiv(ACT)) AS ERR
    FROM AEW WHERE LAYER=l
    GROUP BY ROW) AS E
  WHERE A.COL=W.ROW AND W.COL=E.COL
    AND A.LAYER=l-2 AND W.LAYER=l-1;
    AND A.ITER=i AND W.ITER=i;
end for
  
--Update the first set of weights (on the inputs)
SELECT i+1, 1, ROW, COL, 
       MAT - matmul(t(ACT), ERR) * 0.00000001
BULK COLLECT INTO W 
FROM AEW WHERE LAYER=1;
\end{SQL}
%\vspace{-10 pt}
 \caption{SQL code to implement the backward pass for iteration $i$ of a feed-forward deep network with eight hidden layers.
}
 \label{fig:backprop}
\vspace{-15 pt}
\end{figure}
 
\subsection{So, What's the Catch?}
\label{sec:problem}
In writing a loop, the SQL programmer used a database table to pass state between iterations.
In our example, this is done by utilizing the \texttt{AEW} table, which stores the error being back-propagated through each
of the connections from layer $l + 1$ to layer $l$ in the network, for each of the data points in the current learning batch.  
If there are 100,000 neurons in two adjacent layers  in a fully-connected network and 1,000 data points in a batch, 
then there are $(100,000)^2$ such connections for each of the 1,000 data points, or $10^{13}$ values
stored in all.  Using single-precision floating point value, a debilitating 40TB of data must be materialized.

Storing the set of per-connection errors is a very intuitive choice as a way to communicate among loops iterations, especially
since the per-connection errors are subsequently aggregated in two ways (one to compute the new weights at a layer, and one to
compute the new set of per-connection errors passed to the next layer). 
But forcing the system to materialize this table can result
in a very inefficient computation.  
This \emph{could} be implemented by pipelining the computation creating the new
data for the \texttt{AEW}
table directly into the two subsequent aggregations, but this possibility has been lost when the programmer  
asked that the new data be \texttt{BULK} \texttt{COLLECT}ed into \texttt{AEW}.  

Note that this
is not merely a case of a poor choice on the part of the programmer.  In order to write a loop, state has to be passed
from one iteration to another, and it is this state that made it impossible for the system to realize an ideal implementation.
This is the pitfall of imperative---rather than declarative---programming.

\section{Extensions to SQL}

In this section, we consider a couple of extensions to SQL that make it possible for a programmer (either a human or a
deep learning tool chain) to declaratively specify recursive computations such as back-propagation, without control flow.

\subsection{The Extensions}

We introduce these SQL extensions in the context of a classic introductory programming problem:
implementing Pascal's triangle, which recursively defines binomial coefficients.  Specifically, the goal is to
build a matrix such that the entry in row $i$ and column $j$ is $\binom{i}{j}$ (or $i$ choose $j$).  The triangle is defined recursively so that
for any integers $i\geq0$ and $j\in[1,i-1]$, $\binom{i}{j}=\binom{i-1}{j-1}+\binom{i-1}{j}$:
\begin{center}
\begin{tabular}{>{$}l<{$}|*{7}{c}}
\multicolumn{1}{l}{$i$} &&&&&&&\\\cline{1-1} 
0 &1&&&&&&\\
1 &1&1&&&&&\\
2 &1&2&1&&&&\\
3 &1&3&3&1&&&\\
4 &1&4&6&4&1&&\\\hline
\multicolumn{1}{l}{} &0&1&2&3&4\\\cline{2-8}
\multicolumn{1}{l}{} &\multicolumn{7}{c}{$j$}
\end{tabular}
\end{center}
Our extended SQL allows for multiple versions of a database table; versions are accessed via \textit{array-style indices}.  For example, we can
define a database table storing the binomial coefficient $\binom{0}{0}$ as:

\begin{SQL}
CREATE TABLE pascalsTri[0][0] (val) AS
  SELECT val FROM VALUES (1);
\end{SQL}

\noindent The table \texttt{pascalsTri[0][0]} can now be queried like any other database table, and various versions of the tables can be defined
recursively.  For example, we can define all of the cases where $j = i$ (the diagonal of the triangle) as:

\begin{SQL}
CREATE TABLE pascalsTri[i:1...][i] (val) AS
  SELECT * FROM pascalsTri[i-1][i-1];
\end{SQL}

\noindent And all of the cases where $j = 0$ as:

\begin{SQL}
CREATE TABLE pascalsTri[i:1...][0] (val) AS
  SELECT * FROM pascalsTri[i-1][0];
\end{SQL}

\noindent Finally, we can fill in the rest of the cells in the triangle via one more recursive relationship:

\begin{SQL}
CREATE TABLE pascalsTri[i:2...][j:1...i-1](val) AS
  SELECT pt1.val + pt2.val AS val
  FROM pascalsTri[i-1][j-1] AS pt1, 
       pascalsTri[i-1][j] AS pt2;
\end{SQL}

\noindent Note that this differs quite a bit from classical, recursive SQL, where the goal is typically to compute a fix-point of a set.  Here, there
is no fix-point computation.  In fact, this particular recurrence defines an infinite number of versions of the \texttt{pascalsTri} table.  
Since there can be an infinite number of such tables, those tables are materialized on-demand.  A programmer can issue the query:

\begin{SQL}
SELECT * FROM pascalsTri[56][23];
\end{SQL}

\noindent In which case the system will unwind the recursion, writing the required computation as a single relational algebra statement.  A programmer
may ask questions about multiple versions of a table at the same time (without having each one be computed separately):

\begin{SQL}
EXECUTE (
  FOR j IN 0...50:
    SELECT * FROM pascalsTri[50][j]);
\end{SQL}

\noindent
By definition, all of the queries/statements within an \texttt{EXECUTE} command are executed as part of the same query plan.
Thus, this would be compiled into a single relational algebra statement that produces all 51 of the requested tables, 
under the constraint that each of those 51 tables must be materialized (without such a constraint, the resulting physical execution plan may pipeline
one or more of those tables, so that they exist only ephemerally and cannot be returned as a query result).
If a programmer wished to materialize all of these tables so that they could be used subsequently without re-computation, s/he could use:

\begin{SQL}
EXECUTE (
  FOR j IN 0...50:
    MATERIALIZE pascalsTri[50][j]);
\end{SQL}

\noindent which materializes the tables for later use.
Finally, we introduce a multi-table \texttt{UNION} operator that merges multiple, recursively-defined tables.  This makes it possible
to define recursive relationships that span multiple tables.  For example, a series of tables storing the various Fibonacci numbers (where $Fib(i) = 
Fib(i - 1) + Fib(i - 2)$ and $Fib(1) = Fib(2) = 1$) can be defined as:

\begin{SQL}
CREATE TABLE Fibonacci[i:0...1] (val) AS
  SELECT * FROM VALUES (1); 

CREATE TABLE Fibonacci[i:2...] (val) AS
  SELECT SUM (VAL) FROM UNION Fibonacci[i-2...i-1];
\end{SQL}

\noindent In general, \texttt{UNION} can be used to combine various subsets of recursively defined tables.  For example, one could refer to
\texttt{UNION pascalsTri[i:0...50][0...i]} which would flatten the first 51 rows of Pascal's triangle into a single multiset.

\vspace{10 pt}
\subsection{Learning Using Recursive SQL}

\begin{figure}[t]
\begin{center}
\includegraphics[width=2.4in]{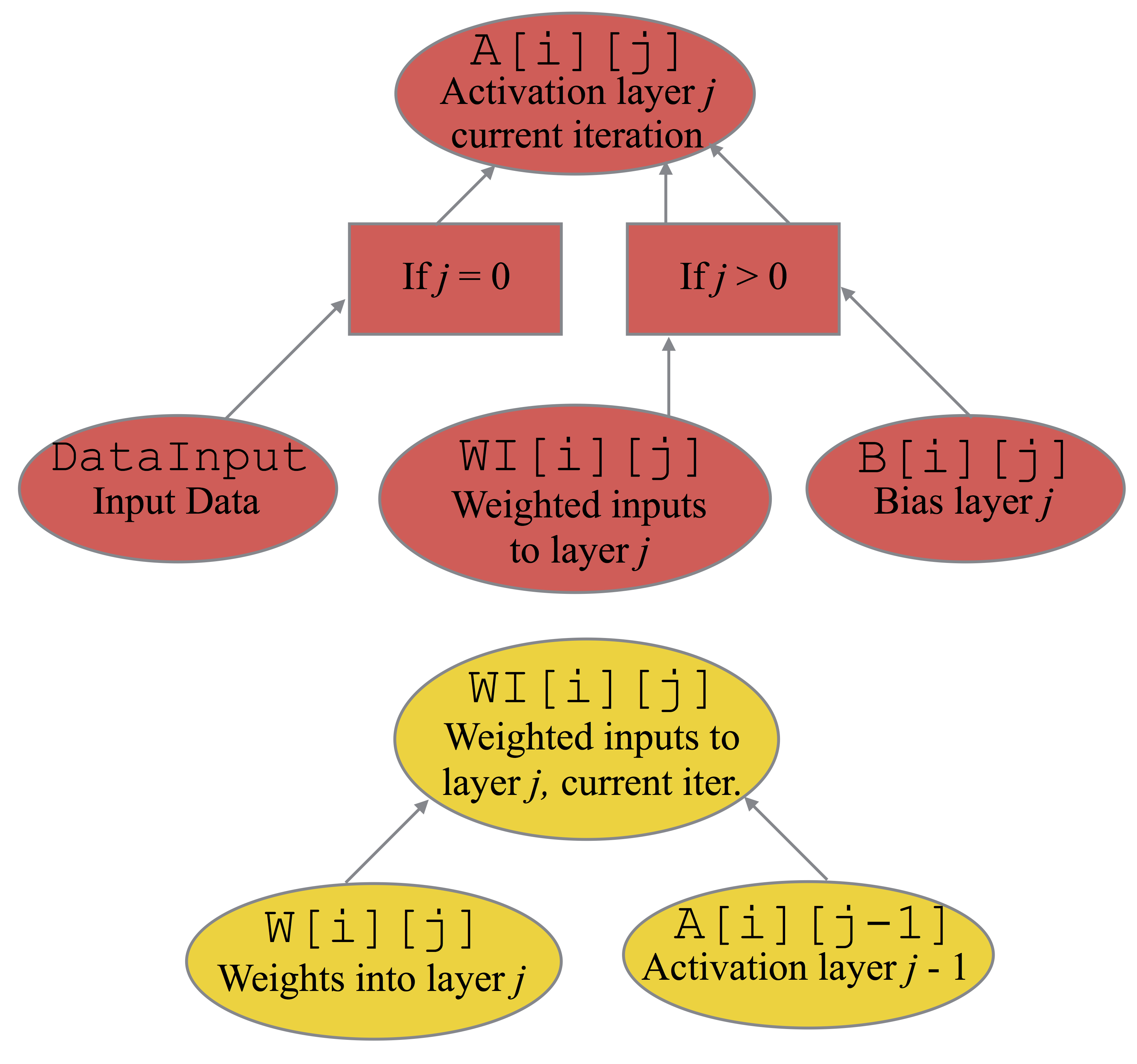}
\end{center}
\vspace{-5 pt}
 \caption{Dependencies in the forward pass through nine layers of SQL-based NN learning.}
 \label{fig:foward-ffnn}
\end{figure}

{\color{black} With our SQL extensions, we can rewrite the 
aforementioned forward-backward passes to eliminate imperative control flow by declaratively expressing the
various dependencies among the activations, weights, and errors.  

\vspace{5 pt}
\noindent
\textbf{Forward pass.} The forward pass is concerned with computing the level of activation of the neurons
at each layer.  The activations of all neurons in layer $j$ at learning iteration $i$ 
are given in table \texttt{A[i][j]}.  Activations are computed using the
weighted sum of the outputs of all of the neurons at the last level; the weighted sums for the layer
$j$ at learning iteration $i$ is given in the table \texttt{WI[i][j]}.
The dependencies making up the forward pass
are depicted in Figure \ref{fig:foward-ffnn}.
The corresponding SQL code is as follows.}  The forward pass begins by loading the first layer of
activations with the input data:

\begin{SQL}
CREATE TABLE A[i:0...][j:0](COL, ACT) AS
  SELECT DI.COL, DI.VAL
  FROM DATA_INPUT AS DI;
\end{SQL}

\vspace{10 pt}
\noindent We then send the activation across the links in the network:

\begin{SQL}
CREATE TABLE WI[i:0...][j:1...9](COL, VAL) AS
  SELECT W.COL, SUM(matmul(A.ACT, W.MAT))
  FROM W[i][j] AS w, A[i][j-1] AS A
  WHERE W.ROW = A.COL
  GROUP BY W.COL;
\end{SQL}

\noindent Those links are then used to compute subsequent activations:

\begin{SQL}
CREATE TABLE A[i:0...][j:1...8](COL, ACT) AS
  SELECT WI.COL, relu(WI.VAL + B.VEC)
  FROM WI[i][j] AS WI, B[i][j] AS B
  WHERE WI.COL = B.COL;
\end{SQL}

\noindent And finally used to perform the prediction:

\begin{SQL}
CREATE TABLE A[i:0...][j:9](COL, ACT) AS
  SELECT WI.COL, softmax(WI.VAL + B.VEC)
  FROM WI[i][j] AS WI, B[i][j] AS B
  WHERE WI.COL = B.COL;
\end{SQL}

\begin{figure}[t]
\begin{center}
\includegraphics[width=3.0in]{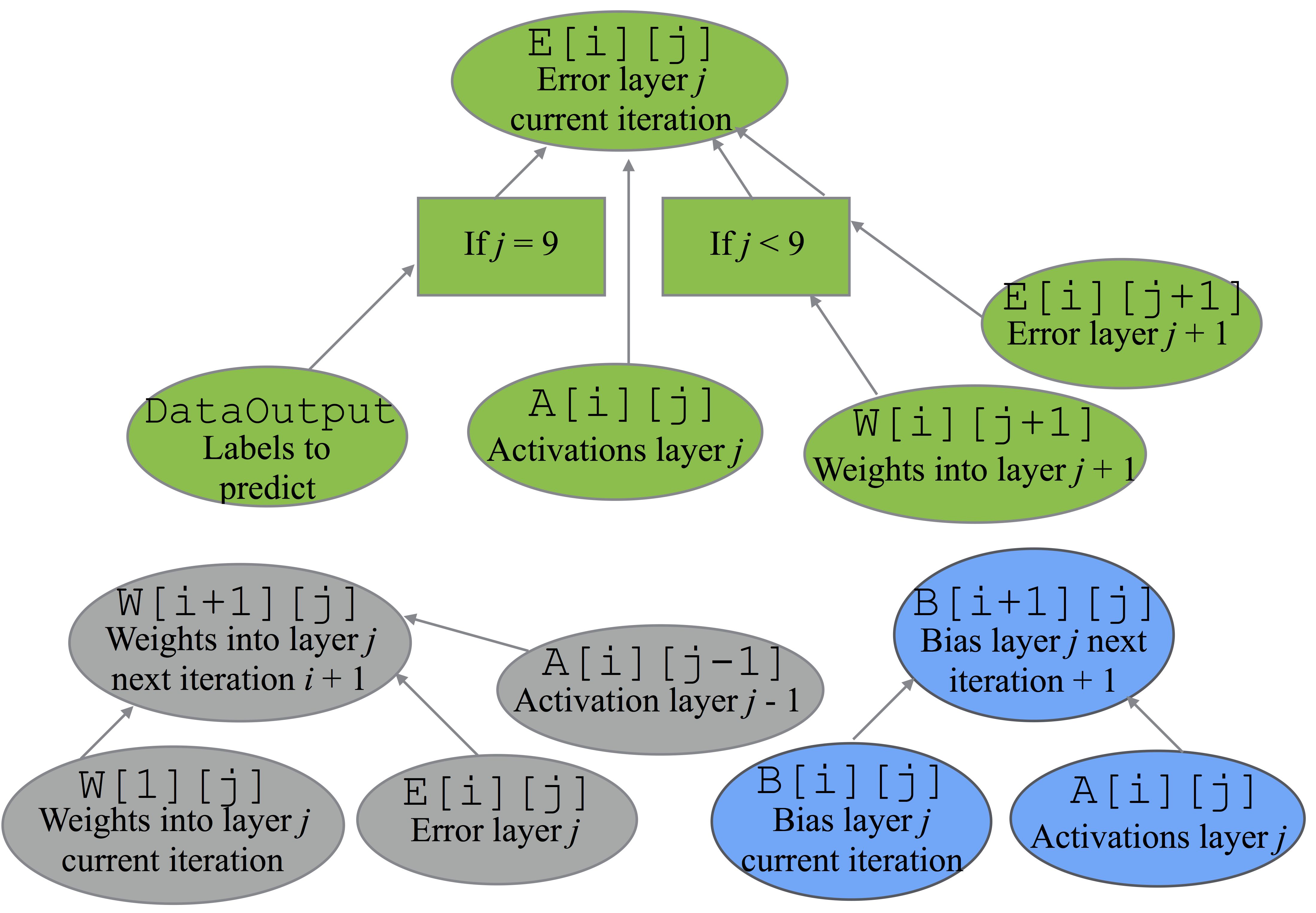}
\end{center}
\vspace{-5 pt}
 \caption{Dependencies in the backward pass of SQL-based NN learning.}
 \label{fig:backward-ffnn}
\end{figure}

{\color{black}
\vspace{5 pt}
\noindent
\textbf{Backward pass.} In the backward pass, the errors are pushed backward through the network.  
The error being pushed through layer $j$ in learning iteration $i$ are stored in the table
\texttt{E[i][j]}.  These errors are used to update all of the network's weights (the weights directly affecting
layer $j$ in learning iteration $i$ are stored in \texttt{W[i][j]}) as well as biases (stored
in \texttt{B[i][j]}). 
The recursive dependencies making up the backward pass
are shown in Figure \ref{fig:backward-ffnn}.
We begin the SQL code for the backward pass with the initialization of the error:}

\begin{SQL}
CREATE TABLE E[i:0...][j:9](COL, ERR) AS
  SELECT A.COL, crossentropyderiv(A.ACT, DO.VAL)
  FROM A[i][j] AS A, DATA_OUTPUT AS DO;
\end{SQL}

\noindent 
At subsequent layers, the error is:
  
\begin{SQL}    
CREATE TABLE E[i:0...][j:1...8](COL, ERR) AS
  SELECT W.ROW, SUM(matmul(E.ERR, t(W.MAT))
                    * reluderiv(A.ACT))
  FROM A[i][j] AS A, E[i][j+1] AS E,
       W[i][j+1] AS W
  WHERE A.COL = W.ROW AND W.COL = E.COL
  GROUP BY W.ROW;
\end{SQL}

\noindent Now we use the error to update the weights:

\begin{SQL}  
CREATE TABLE W[i:1...][j:1...9](ROW, COL, MAT) AS
  SELECT W.ROW, W.COL, 
     W.MAT - matmul(t(A.ACT), E.ERR) * 0.00000001
  FROM W[i-1][j] AS W, E[i-1][j] AS E,
       A[i-1][j-1] AS A
  WHERE A.COL = W.ROW AND W.COL = E.COL;
\end{SQL}

\noindent And the biases:

\begin{SQL}  
CREATE TABLE B[i:1...][j:1...9](COL, VEC) AS
  SELECT B.COL, 
       B.VEC - reducebyrow(E.ERR) * 0.00000001
  FROM B[i-1][j] AS B, E[i-1][j] AS E
  WHERE B.COL = E.COL;
\end{SQL}

We now have a fully declarative implementation of neural network learning.

\section{Executing Recursive Plans}

The recursive specifications of the last section address the problem of how to succinctly and declaratively specify complicated recursive
computations. Yet the question remains: How can the very large and complex computations associated with such specifications be compiled and executed by an RDBMS without
significant modification to the system?

\subsection{Frame-Based Execution}

Our idea for compiling and executing computations written recursively in this fashion is to first compile the recursive computation into a single monolithic relational algebra DAG, and then partition the computation into \emph{frames}, or sub-plans.  Those frames are then optimized and executed independently, with intermediate tables materialized to facilitate communication between frames.  

Frame-based computation
is attractive because if each frame is small enough that an existing query optimizer and execution engine can
handle the frame, the RDBMS optimizer and engine need not be modified in any way.  Further, this iterative execution results in an engine that resembles engines that perform re-optimization during runtime \cite{kabra1998efficient}, 
in the sense that frames are optimized and executed once all of their inputs have been materialized.  Accurate statistics can be collected on those inputs---specifically, the number of distinct attribute values can be collected 
using an algorithm like Alon-Matias-Szegedy \cite{alon1996space}---meaning
that classical problems associated with size estimation errors propagating through a query plan can be avoided.

{\color{black}
\subsection{Heuristic vs. Full Unrolling}

One could imagine two alternatives for implementing a frame-based strategy.  The first is to rely on a heuristic,
such as choosing the outer-most loop index, 
breaking the computation into frames using that index, and so on.  
However, there are several problems with this approach.  First off, we are back to the problem described
in Section~\ref{sec:problem}, where we are choosing to materialize tables in an ad-hoc and potentially dangerous
way (we may materialize a multi-terabyte table). 
Second, we cannot control the size of the frame. 
Too many operations in one frame can mean that the system is unable to optimize and execute that frame, while too few can mean
a poor physical plan with too much materialized data.  Third, if we allow
the recursion to go up as well as down, or skip index values, this will not work.

Instead, we opt for an approach that performs a full unrolling of the recursive computation and turns it to a single, monolithic
computation, and then we define an
optimization problem that attempts to split the computation into frames so as to minimize the likelihood of
materializing a large number of tables.  

\subsection{Plan Unrolling}

Our unrolling algorithm attempts to leverage an existing RDBMS query compiler to transform an
SQL query into an un-optimized relational algebra (RA) plan.
At a high level, the algorithm recursively chases the original query's dependencies.  Whenever a
dependency is found, a lookup table (called the \emph{sub-plan lookup table})
is checked to see if the dependence had previously been compiled.  
If not, the recursive dependency is expanded.  This proceeds until table definitions are reached with no further 
un-compiled
dependencies.  At this point, the recursion unwinds, and any remaining dependencies are recursively expanded.
Eventually, a directed, acyclic graph of RA operations is produced.

All of this is best illustrated by continuing the Pascal's triangle example.  
Assume that a programmer asks for the following:

\begin{SQL}
SELECT * FROM pascalsTri[3][2] (val);
\end{SQL}

The unrolling algorithm begins by analyzing the recursive SQL table definitions, 
and determining which tables this query depends on.
Since this query is covered by the definition

\begin{SQL}
CREATE TABLE pascalsTri[i:2...][j:1...i-1](val);
\end{SQL}

\noindent
which depends upon \texttt{pascalsTri[i-1][j-1]} (evaluating to \texttt{pascalsTri[2][1]}) and
\texttt{pascalsTri[i-1][j]} (evaluatining to \texttt{pascalsTri[2][2]}, we must determine the dependencies
for \texttt{pascalsTri[2][1]} and \texttt{pascalsTri[2][2]}.  The latter is covered by the definition

\begin{SQL}
CREATE TABLE pascalsTri[i:1...][i](val);
\end{SQL}

\noindent
This definition in turn depends upon \texttt{pascalsTri[i-1][i-1]}.  The expression evaluates to \texttt{pascalsTri[1][1]}
which depends upon \texttt{pascalsTri[0][0]}.  Since
\texttt{pascalsTri[0][0]} is defined directly as:

\begin{SQL}
SELECT val FROM VALUES (1);
\end{SQL}

\noindent
the recursion bottoms out, and this %SQL (I comment this word mainly to fix an overflow at the end of this sentence) 
query is compiled (using the existing compiler) into an RA
plan.  The root of this RA is inserted into the sub-plan lookup table, with key \texttt{pascalsTri[0][0]}.

The recursion then unwinds to \texttt{pascalsTri[1][1]}.
Since all of this table's dependencies 
are now covered, we are ready to compile \texttt{pascalsTri[1][1]}'s SQL into RA.  We textually replace 
\texttt{pascalsTri[i-1][i-1]} in the definition of \texttt{CREATE TABLE pascalsTri[i:1...][i](val)} with
the dummy table \texttt{pascalsTri\_0\_0} to obtain:

\begin{SQL}
SELECT * FROM pascalsTri_0_0;
\end{SQL}

\noindent
and then compile this into RA.  The scan of \texttt{pascalsTri\_0\_0} in the resulting RA
is replaced with a link from the root node of the RA for \texttt{pascalsTri[0][0]} (obtained from the 
sub-plan lookup table), and we now have a complete RA for \texttt{pascalsTri[1][1]}, which is also put into
the sub-plan lookup table.

The recursion unwinds to \texttt{pascalsTri[2][2]}, which likewise is obtained by compiling:

\begin{SQL}
SELECT * FROM pascalsTri_1_1;
\end{SQL}

\noindent
The %table (comment this to remove overflow of \texttt{pascalsTri[1][1]}, Jia)
scan of \texttt{pascalsTri\_1\_1} in the resulting RA
is replaced with a link from the root of the RA for \texttt{pascalsTri[1][1]}, and we now have a complete
plan for \texttt{pascalsTri[2][2]}.

The recursion then unwinds to 
\texttt{pascalsTri[3][2]} which depends upon both \texttt{pascalsTri[2][2]} (now present in the sub-plan lookup table),
and \texttt{pascalsTri[2][1]}.  Since the latter is not present in the lookup table, we must recursively chase its
dependencies.  Once this recursion unwinds, we are ready to compile the SQL for \texttt{pascalsTri[3][2]}:

\begin{SQL}
SELECT pt1.val + pt2.val AS val
FROM pascalsTri_2_1 AS pt1,
     pascalsTri_2_2 AS pt2;
\end{SQL}

Replacing the table scans in the resulting RA with links from the RA plans for \texttt{pascalsTri[2][2]} and
\texttt{pascalsTri[2][1]} completes the compilation into a single, monolithic plan. }

\section{Plan Decomposition}

The algorithm of the previous section
produces a monolithic plan.
We now consider the problem of computing the best cut of a very large plan into a set of frames.

\subsection{Intuition}

The cost incurred when utilizing frames is twofold.  First, it restricts 
the ability
of the system's logical and physical optimizer to find optimization opportunities.  For example, if the logical plan
$((R \Join S) \Join T)$ is optimal but the input plan $((R \Join T) \Join S)$ is cut into frames $f_1 = (R \Join T)$ and
$f_2 = (f_1 \Join S)$, it is impossible to realize this optimal plan.
In practice, we address this by placing a minimum size on frames as {\color{black} larger frames make it more likely that high-quality join
orderings will still be present in the frame.}

More significant is the requirement that the contents of already-executed frames be saved so that later frames
may utilize them.  This can introduce significant I/O compared to a monolithic execution.
Thus we may attempt to cut into frames to minimize the number of bytes traveling over cut edges.
Unfortunately, this is unreasonable as it is well-understood that estimation
errors propagate through a plan; in the upper reaches of a huge plan, it is going to be impossible to estimate
the number of bytes traveling over edges.

Instead, we find that spitting the plan into frames so as to reduce the number of \emph{pipeline breakers} induced is a reasonable goal.  A pipeline breaker occurs when the output of one operator must be materialized to disk or
transferred over the network, as opposed to being directly communicated from operator to operator via CPU cache, or, in the worst
case, via RAM.  An induced pipeline breaker is one that would not have been present in an optimal physical plan but was forced by the cut.

\subsection{Quadratic Assignment Formulation}

{\color{black} Given a query plan, it is
unclear whether a cut that separates two operators into different frames 
will induce a pipeline breaker.}  We model this uncertainty
using probability and seek to minimize the expected number of pipeline breakers induced by the set of chosen frames.

This is ``probability'' in the Bayesian rather than frequentist
sense, in that it represents a level of certainty
or belief in the pipelineability of various operators.  
For the $i$th and $j$th operators in the query plan,
let $N_{ij}$ be a random variable that takes the value $1$ if operator $i$ is pipelined into operator $j$ were
the entire plan optimized and executed as a unit, and $0$ otherwise.

Let the query plan to be cut into frames be represented as a directed graph having $n$ vertices, represented as a binary matrix
\textbf{E},
where $e_{ij}$ is one (that is, there is an edge from
vertex $i$ to vertex $j$) if the output of operator $i$ is directly
consumed by operator $j$. $e_{ij}$ is zero otherwise.
We would like to split the graph into $m$ frames. 
We define the \emph{split} of a query plan to be a matrix $\textbf{X}=(x_{ij})_{n \times n}$, 
where each row would be one frame so that $x_{ij}=1$ if operator $i$ is in a different frame from operator $j$ (that is, they 
have been cut apart)
and 0 otherwise. 
Given this, the goal is to minimize:
\begin{align} cost(\textbf{X}) &= E\left[\sum_{i=1}^{n} \sum_{j=1}^n e_{ij} x_{ij} N_{ij} \right] 
    &= \sum_{i=1}^{n} \sum_{j=1}^n e_{ij} x_{ij} E\left[ N_{ij} \right] \nonumber
\end{align}
This computes the expected number of pipeline breakers induced, as for us to induce a new pipeline breaker via the 
cut, (a) operator $j$ must consume the output from operator $i$, (b) operator $i$ and $j$ must be separated by the cut, and
(c) operator $i$ should have been pipelined into operator $j$ in the optimal execution.

We can re-write the objective function by instead letting the matrix $\textbf{X}=(x_{ij})_{n \times m}$ be an \emph{assignment matrix},
where $\sum_i x_{ij} = 1$, and each $x_{ij}$ is either one or zero.  Then, $x_{ij}$ is one if operator $i$ is put into frame $j$ and
we have:

\begin{align} 
cost(\textbf{X}) =
&\left( \sum_{i=1}^{n} \sum_{j=1}^n \sum_{a=1}^{m} \sum_{b=1}^m e_{ij} x_{ia} x_{jb} E\left[ N_{ij} \right] \right) - \nonumber \\
&\left( \sum_{i=1}^{n} \sum_{j=1}^n \sum_{a=1}^{m} e_{ij} x_{ia} x_{ja} E\left[ N_{ij} \right] \right) \nonumber
\end{align}

Letting $c_{ijab} = e_{ij} E\left[ N_{ij} \right] - \delta_{ab} e_{ij} E\left[ N_{ij} \right]$ $=$ 
$e_{ij} E\left[ N_{ij} \right] (1 - \delta_{ab})$ where $\delta_{ab}$ is the Kronecker delta
function, we then have:

\begin{align}
cost(\textbf{X}) =
\sum_{i=1}^{n} \sum_{j=1}^n \sum_{a=1}^{m} \sum_{b=1}^m c_{ijab} x_{ia} x_{jb} \nonumber
\end{align}

The trivial solution to choosing $\textbf{X}$ to minimize this cost function is to put all or most operators in the same frame, 
but that would result in a query plan that is not split in a meaningful way. Therefore we need to add a constraint on the upper bound of operators in each frame: $min \leq \sum_j x_{ij} \leq max$ for some maximum frame size.

The resulting optimization problem is not novel: it is an instance of the problem popularly
known as the \emph{generalized quadratic assignment problem},
or GQAP \cite{lee:2004}, where the goal is to map tasks or machinery (in this case, the various operations we are executing) into locations or
facilities (in this case, the various frames).  GQAP generalizes the classical quadratic assignment problem by allowing multiple tasks or
pieces of machinery to be mapped into the same location or facility (in the classical formulation, only one task is allowed per facility).  
Unfortunately, both GQAP and classical quadratic assignment are NP-hard, and inapproximable.  

In our instance of the problem, we actually have one additional constraint that is not expressible within the standard GQAP framework.
A simple minimization of the objective function could
result in a sequence of frames that may not be executable because they contain
circular dependencies. In order to ensure that we have no circular dependencies, 
we have to make the intermediate value that a frame uses available before it is executed. To do this, we take the natural ordering
of the frames to be meaningful, in the sense that frame $a$ is executed before frame $b$ when $a < b$, and for each edge 
$e_{ij}$ in the computational graph, we introduce the constraint that 
for $a, b$ where $x_{ia} = 1$ and $x_{jb} = 1$, it must be the case that $a \leq b$.  

\vspace{15 pt}
\subsection{Cost Model}
\label{sec:cost-model}

So far, we have not discussed the precise nature of the various $N_{ij}$ variables that control whether the output of operator $i$ is
pipelined into operator $j$ in a single, uncut, optimized and executed version of the computation.  Specifically, we need to compute
the value of $E\left[ N_{ij} \right]$ required by our GQAP instance.  Since each $N_{ij}$ is a binary variable, $E\left[ N_{ij} \right]$
is simply the probability that $N_{ij}$ evaluates to one.  Let $p_{ij}$ denote this probability.
In keeping with our Bayesian view, we define the various $p_{ij}$ values as follows:

\begin{itemize}
\item If the output of operator $i$ has one single consumer (operator $j$) and operator $j$ is a selection or an aggregation, then 
$p_{ij}$ is 1. The reason for this is that in the system we are building on (SimSQL \cite{cai2013simulation}), 
it is always possible to pipeline into a selection or an aggregation.  Selections are always pipelineable, and in SimSQL, if operator
$j$ is an aggregation, then a corresponding pre-aggregation will be added to the end of the pipeline executing operation $j$.  
This pre-aggregation maintains a hash table for each group encountered in the aggregation, and as new data are encountered, statistics
for each data object are added to the corresponding group.  As long as the number of groups is small and the summary statistics compact,
this can radically reduce the amount of data that needs to be shuffled to implement the aggregation.

\item If the output of operator $i$ has one single consumer (operator $j$) but operator $j$ is not a selection or an aggregation,
then $p_{ij}$ is estimated using past workloads.  That is, based off of workload history,
we compute the fraction of the time that operator $i$'s type of operation
is pipelined into the type of operator $j$'s operation, and use that for $p_{ij}$.

\item In SimSQL, if operator $i$ has multiple consumers, then the output of operator $i$ can be pipelined into only one of them (the
output will be saved to disk and then the other operators will be executed subsequently, reading the saved output).  Hence, if 
there are $k$ consumers of operator $i$, and operator $j$ is a selection or an aggregation, then $p_{ij} = \frac{1}{k}$.  Otherwise, 
if, according to workload history, the traction of the time that operator $i$'s type of operation
is pipelined into the type of operator $j$'s operation is $f$, then $p_{ij} = \frac{f}{k}$.
\end{itemize}

\subsection{Heuristic Solution}
\label{sec:heuristic}

Generalized quadratic assignment is a very difficult problem \cite{burkard2013assignment}.  Therefore, we turn to developing
a heuristic solution that may work
for our application.

One simple idea is a greedy algorithm that builds frames, one at a time.  We treat the relational algebra computation as a graph, labeling
each edge with the associated $p_{ij}$ value.
The algorithm starts from a source operator and adds operators to the frame iteratively until the frame size exceeds $min$,
always adding the operator that directly depends on an operator in the current 
frame that would yield the smallest increase in the cost of the whole frame 
(the increase is the cost of the new edge cut, minus the cost of an edge that is now internal to the frame).  In order
to ensure that we do not build frames that
have circular dependencies, when we add a new operation $o_i$ to the frame
where the edge $o_i$ from $o_j$ is present in the graph but $o_j$ is not yet part of any frame, 
we add $o_j$ to the frame.  This may 
in turn trigger the recursive addition of new operators to the frame.

\begin{figure*}

\hspace{0.2cm}
\begin{tabular}{p{0.22\textwidth}p{0.22\textwidth}p{0.22\textwidth}p{0.22\textwidth}}

\begin{minipage}{0.22\textwidth}
\centering
\includegraphics[scale=0.35]{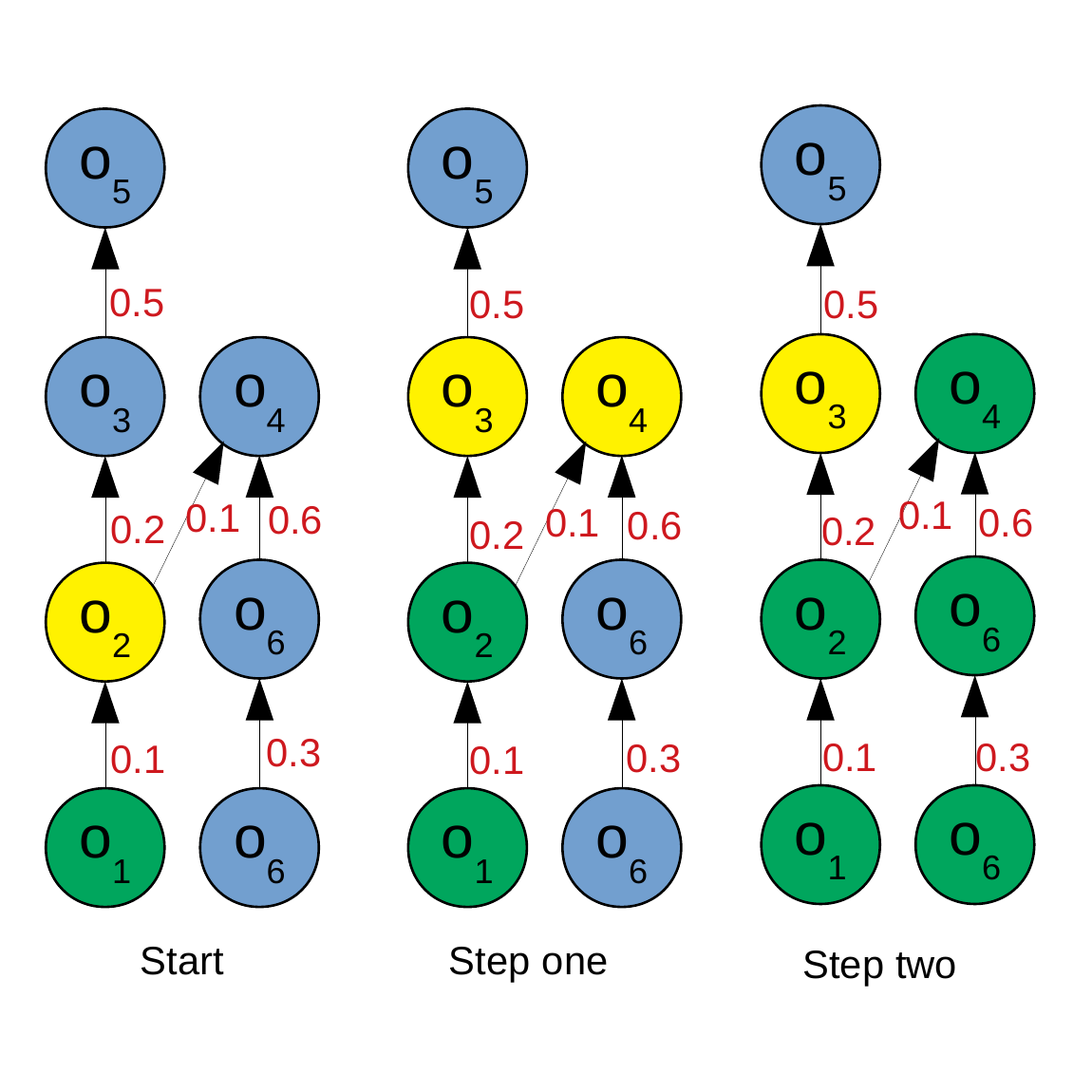}
\end{minipage} &

\begin{minipage}{0.22\textwidth}
\centering
\includegraphics[scale=0.35]{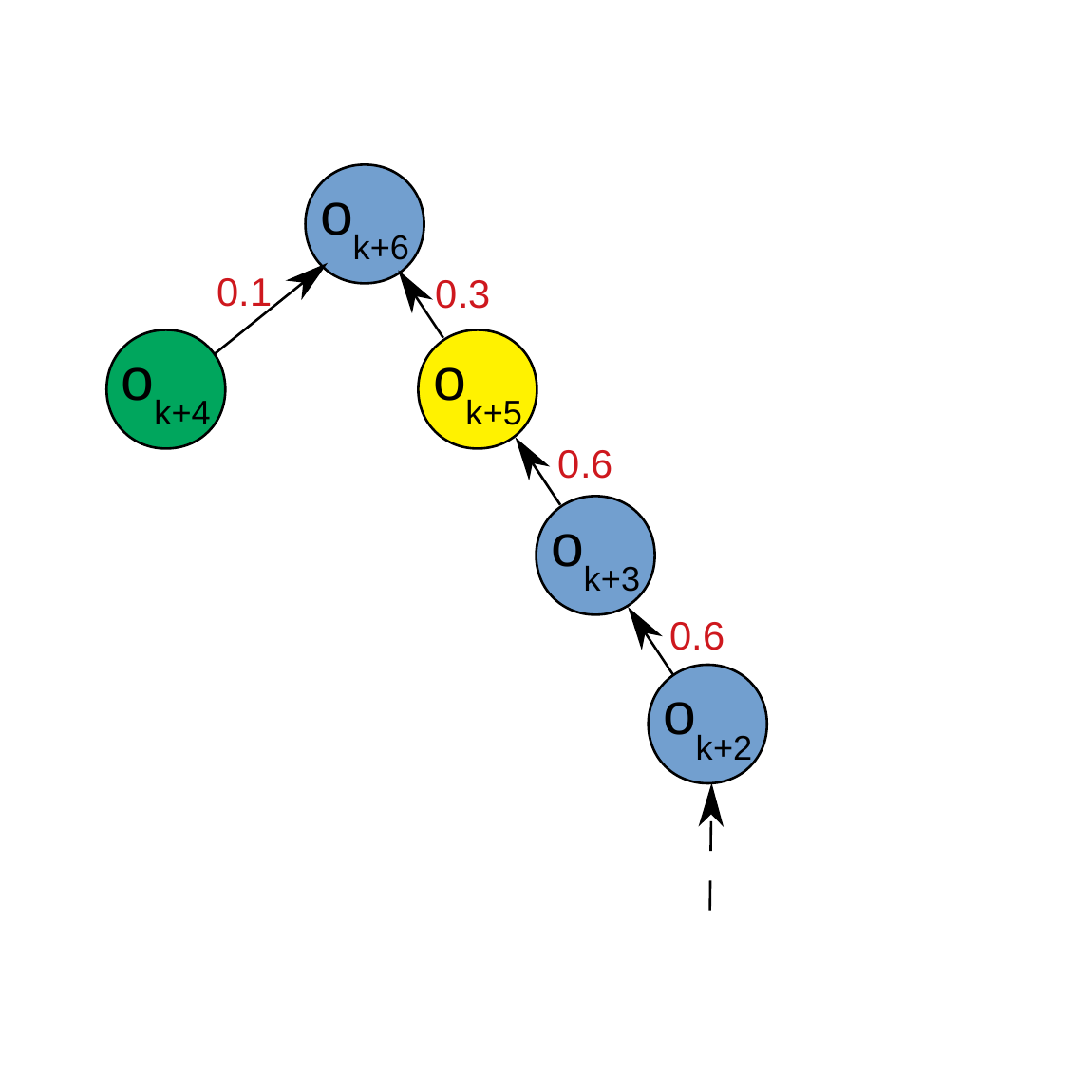}
\end{minipage}&

\begin{minipage}{0.22\textwidth}
\centering
\includegraphics[scale=0.35]{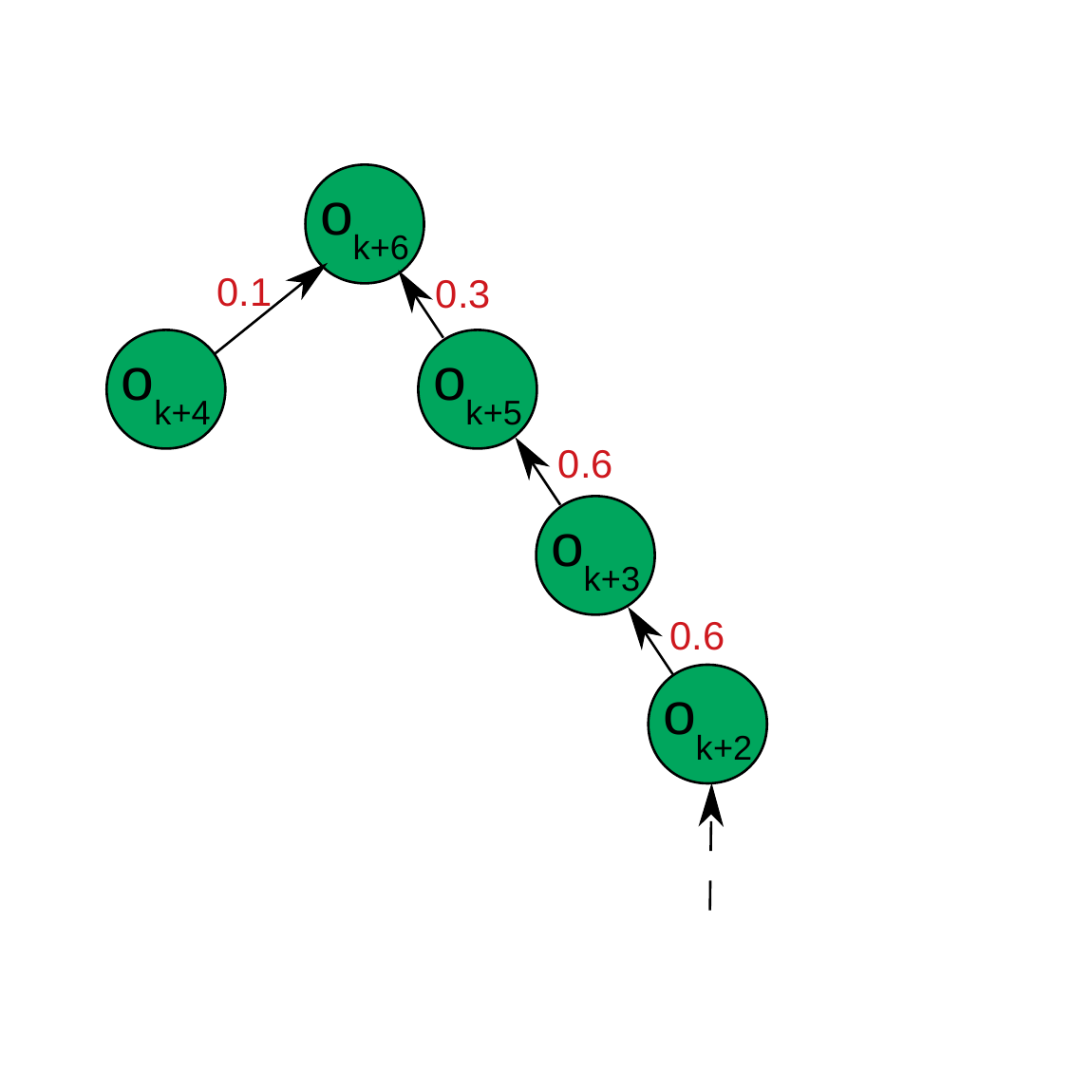}
\end{minipage}&

\begin{minipage}{0.22\textwidth}
\centering
\includegraphics[scale=0.35]{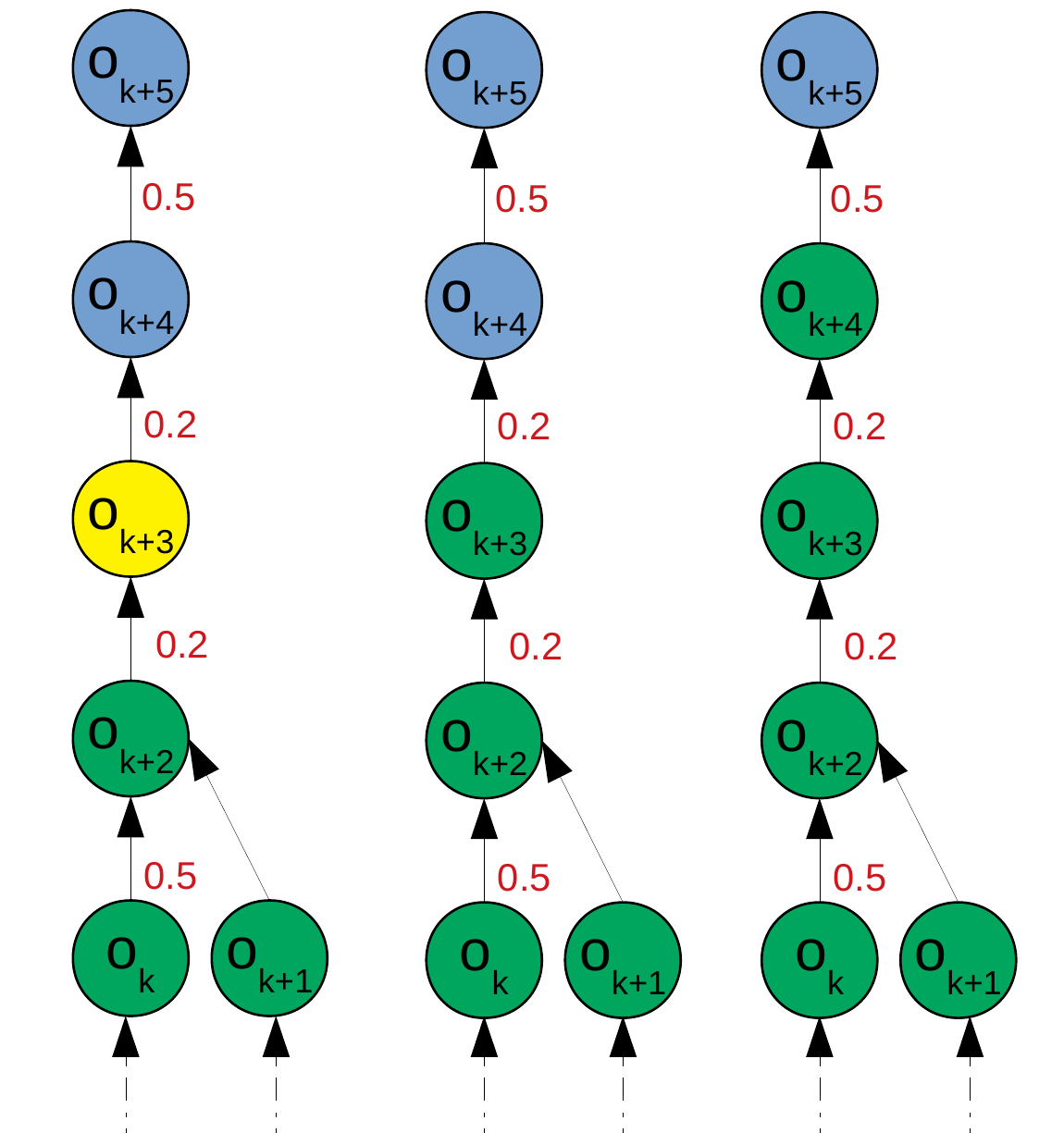}
\end{minipage}
\\
%\vspace{3 pt}

(a) The greedy algorithm. The values in red represent the costs. The operators in green are selected as part
of the frame. The operators in yellow are under consideration. &

(b) Greedy if the source operator is badly chosen. Adding the source operator adds the rest of the graph to the frame. &

(c) Greedy where the upper bound is set to $max=7$.&

(d) Trying three different frame sizes. The middle frame will be selected.\\

\end{tabular}

\vspace{-10 pt}
\caption{Greedily cutting a frame from a compute plan.}
\label{fig:algoextract}
\end{figure*}

An illustration of how the algorithm works is given in Figure \ref{fig:algoextract}(a).  We begin with operation $o_1$, adding
operation $o_2$.  Then we add $o_4$ since it has a lower cost (the cost is $0 - 0.1 = -0.1$) than $o_3$ (cost is $0.5 - 0.3 = 0.2$).  
Since $o_4$ requires that we have computed
$o_6$, we add $o_6$ and all of its un-added dependencies.

%\begin{algorithm}[H]
%\caption{Greedy}\label{euclid}
%\begin{algorithmic}[1]
%\Function{GetCut(sources)}{}
%\State $\textit{$edge\_node\_list$} \gets \text{empty}$
%\State $\textit{$frame$} \gets \text{empty}$
%\State $\textit{$src$} \gets sources.\text{pop()}$
%\State $\textit{$frame$}.\text{add($src$)}$
%\State $\textit{$frame$}.\text{addAll($src$.parents)}$
%		\State \textbf{repeat}: 
%            \State $\textit{$best\_cost$} \gets \text{inf}$
%            \State $\textit{$best\_edge$} \gets \text{nil}$
%
%            \For{$e$ in $edge\_node\_list$}
%%
%                \State $\textit{$new\_frame$} \gets frame$
%%                \State $\textit{$new\_frame$}.\text{add($e$)}$
%%                \State $\textit{$new\_frame$}.\text{addAll(getChildrenRecursively($e$, $frame$))}$%\Comment{Children we need as a result of adding e}
%
%                \State $\textit{$cost$} \gets \text{getCost(frame)}$
%
%                \If {$cost < best\_cost$}
%
%                  \State $\textit{$best\_cost$} \gets cost$
%                  \State $\textit{$best\_edge$} \gets best\_edge$       
%                \EndIf
%            \EndFor
%            \State $\textit{$frame$}.\text{add($best\_edge$)}$
%            \State $\textit{$frame$}\text{.addAll(getChildrenRecursively($best\_edge\_frame$))}$
%		\State \textbf{until} \emph{$frame.size \leq D$}
%        \State \textbf{return} $frame$
%\EndFunction
%\end{algorithmic}
%\end{algorithm}
 
There are some problems with this algorithm.  First, it is highly dependent on the chosen starting point; choosing a 
bad start can lead to a poor cut.  Consider Figure \ref{fig:algoextract}(b).  Operation $o_{k+4}$ is near the end of a very
long computation.  If we choose this operation to start with, we will next add $o_{k+5}$ which will cause the entire query
plan to be recursively added into the frame.
This makes it impossible to keep the number of operators in the frame below $max$.

We can remedy this by running the greedy algorithm repeatedly, starting with each possible operation.  For each run, we begin
recording the frames (and associated costs) that were generated as soon as the frame size exceeds $min$, 
stop recording (and growing) the frames when its size meets or would exceed $max$.  Out of all of the frames generated from 
each possible starting point, we choose the frame with the minimum cost.
This is illustrated in Figure \ref{fig:algoextract}(d), with a lower
bound of three and an upper bound of five.  In this case, the frame of size four is chosen.

There is a natural concern that a high-cost edge may block the 
discovery of an optimal cut.  For example, we may be at operator $o_1$; we can choose to add operator $o_2$ or operator $o_3$ to the current frame.
Operator $o_3$ has a higher cost; we choose to add $o_2$.  It may be, however, that $o_3$ has a very low-cost link to operator $o_4$ that
we will not discover because we will never add $o_3$.  This can be handled by adding a lookahead to the greedy algorithm.  We have experimented
with this a bit and found that in this particular domain, a purely greedy algorithm seems to do as well as an algorithm with a small
lookahead.

\section{Experiments}

\subsection{Overview}

In this section, we detail a set of experiments aimed at answering the following questions:

\vspace{5 pt}
\noindent
\textit{Can the ideas described in this paper be used to re-purpose an RDBMS so that it can be used
to implement scalable, performant, model parallel ML
computations?}

\vspace{5 pt}
\noindent
We implement the ideas in this paper on top of SimSQL, a research-prototype, distributed
database system \cite{cai2013simulation}.  
SimSQL has a cost-based optimizer, an assortment of implementations 
of the standard relational operations, the ability to pipeline those operations and
make use of ``interesting'' physical data organizations.  It also
has native matrix and vector support \cite{luo2017scalable}.  

Our benchmarking considers
distributed implementations of three ML algorithms:
(1) a multi-layer feed-forward neural network (FF-NN), (2) the
Word2Vec algorithm \cite{Mikolov2013EfficientEO} for learning embeddings of text into a high-dimensional space, 
and (3) a distributed, collapsed Gibbs sampler for LDA \cite{Blei2003LatentDA} (a standard text mining model).
{\color{black} All are widely-used algorithms, and all are quite different. The FFNN is chosen as an ideal case 
for an ML platform such as TensorFlow that is built around GPU support, 
as it consists mostly of matrix operations that run well on a GPU.
Word2Vec is chosen because it naturally requires a huge model.  LDA is interesting
because it benefits the most from a model-parallel implementation.

For the first two neural learners, we compare our RDBMS implementations with the data parallel feed-forward and
Word2Vec implementations that are shipped with 
TensorFlow.  
For the collapsed LDA sampler, we compare with bespoke implementations on top of TensorFlow and Spark.

\vspace{5 pt}
\noindent
\textbf{Scope of Evaluation.} We stress that this is not a ``which system is faster?'' comparison.
SimSQL is implemented in Java and runs on top
of Hadoop MapReduce, with the high latency
that implies.  Hence a platform such as Tensorflow is
likely to be considerably faster than
SimSQL, at least
for learning smaller models (when SimSQL's high fixed costs dominate).}

Rather than determining which system is faster, the specific goal is to study
whether an RDBMS-based, model-parallel learner may be 
a viable alternative to a system such as TensorFlow, and whether it has 
any obvious advantages.

\vspace{5 pt}
\noindent
\textbf{Experimental Details.}
In all of our experiments, all implementations
run the same algorithms over the same data.
Thus, a configuration that runs each iteration 50\% faster than another configuration will reach a given 
target loss value (or log-likelihood) 50\% faster.
Hence, rather than reporting loss values (or log-likelihoods) 
we report per-iteration running times.

All implementations are fully synchronous, for an apples-to-apples comparison. We choose synchronous learning as
there is strong evidence that synchronous learning for large, dense problems
is the most efficient choice \cite{chen2016revisiting, goyal2017accurate}.

{\color{black}
There were two sets of FFNN experiments.  In the first set, EC2 \texttt{r5d.}\-\texttt{2xlarge} CPU machines with 8 cores and 64GB of RAM were used.  In the second set, at various cost levels, we chose
sets of machines to achieve the best performance.  For TensorFlow, this was realized by GPU machines (CPU for parameters); 
for SimSQL, both CPU and GPU machines achieved similar performance.

Word2Vec and LDA were run on clusters of Amazon EC2 \texttt{m2.4}\-\texttt{xlarge} CPU machines, 
each with eight cores and 68GB of RAM. 
GPUs were not used as they are ineffective for these problems---LDA is not a neural learning problem, and Word2Vec's
running time (on TensorFlow) is dominated by parameter server requests, rather than by computations.}

\subsection{Learning Algorithms}

In this subsection, we describe the three different learning algorithms used in the benchmarking.

\noindent
\textbf{(1) A Feed-Forward Neural Network.} 
Our RDBMS-based implementation has already been described extensively.  We use the data parallel, synchronous,
feed-forward network
implementation that ships with TensorFlow as a comparison.  

We use a Wikipedia dump of 4.86 million documents as the input to the feed-forward learner.  The goal is to learn how
to predict the year of the last edit to the article.  There are 17 possible labels in total.  
We pre-process the Wikipedia dump, representing each document as a 60,000-dimensional feature vector, where each
feature corresponds to the number of times a particular unigram or bigram appears in the document.  

In most of our experiments, we use a size 10,000 batch, as recent results have indicated that a relatively large batch
of this size is a reasonable choice for large-scale learning \cite{goyal2017accurate}.

\vspace{5 pt}
\noindent
\textbf{(2) Word2Vec.}  
Word2Vec (W2V) is a two-layer neural network used to generate word embeddings.
We use skip-gram Word2Vec as well as
negative sampling, with 64 negative samples, and
noise contrastive estimation (NCE) loss.
We train our Word2Vec model using the same Wikipedia dump described above, embedding the 1 million most frequent tokens in the 
corpus.
The input and output layers in our experiments both have one million neurons. The neurons of the input layer are connected to the neurons of an intermediate embedding layer, which are further connected to the neurons of the output layer. Therefore, there are two weight matrices of size $10^6 \times d$, where $d$ is the embedding dimensionality. 
The input document is randomly selected and processed with a skip window size of 1. 
On average, each batch has 1240 word pairs.

Our Word2Vec SQL implementation uses three recursive schemas. 
For the weight matrices we use \texttt{weights[i:0...][j:1...2]} 
with attributes \texttt{tokenID} and \texttt{embedVec}. By storing the embedding of each token as a vector, we automatically have a model parallel representation. 
\texttt{embeds[i:0...][j:1...3]} stores the embedding vectors, 
where $j=1$ gives the embeddings corresponding to input labels in a batch, 
$j=2$ gives those corresponding to out labels, 
and $j=3$ gives the negative samples. \texttt{errors[i:0...][j:1...2]} represents the delta updates to be applied back to 
\texttt{weights[i][j]}. 

We compare our RDBMS implementation with the Word2Vec implementation that ships with TensorFlow.

\vspace{5 pt}
\noindent
\textbf{(3) Latent Dirichlet Allocation.}
LDA is a standard text mining algorithm and collapsed Gibbs sampling is a standard learning algorithm for LDA.  The goal
of learning LDA is to learn a set of \emph{topics}, which can identify the words that tend to co-occur with one another.
Collapsed
LDA requires maintaining counts 
of (1) the number of words assigned to each topic in a document, and (2) 
the number of words assigned to each topic in the corpus.  
Workers must repeatedly load a document, cycle through the words in the document, re-assign them to topics,
and update the two sets of counts.
In distributed LDA, since local updates change the global topic counts---and these updates cannot be distributed globally in an
efficient manner---the effect of local updates is typically ignored \cite{Smola:2010} until a synchronization step.  In our LDA
implementation, we divide the input documents into ten subsets.  All of the documents in one subset are processed together.
Later in a synchronized aggregation, the number of words assigned to each topic is updated.  

LDA is also learned over the Wikipedia dump. The dictionary size is $60,000$.

%In the RDBMS, LDA is implemented by annotating the documents with a \texttt{partitionID}. 
%The documents with a \texttt{partitionID = j MOD 10} are grouped into the same partition \texttt{j}, and are processed together. 
In the RDBMS, LDA is implemented by grouping the documents into ten partitions. 
The documents with \texttt{docID/batchSize = j} are assigned to the partition \texttt{j}, and will be processed together. 
The word-to-topic counts for each document are stored in the table \texttt{wordToTopic[i][j](docID,wordID,topicID,\-cnt)},\\
and this table is updated in a per-iteration (\texttt{i}), per-partition (\texttt{j}) manner.
To refer the complete set of topic assignments at the beginning of iteration \texttt{i}, we locally aggregate for the counts in the table \texttt{wordToTopic[i][j]},
and then use an \texttt{UNION} operation to concatenate the aggregated tables. Lastly, a final aggregation is called to get the total topic-word-counts for all documents. 

We build an analogous implementation using Spark resilient distributed datasets (RDDs), as well as on top of 
TensorFlow.
TensorFlow's implementation is ``lightly'' model parallel, in that while data is partitioned, requests to the parameter
server pull only the required portion of the model.
The topic-word counts (\texttt{ntw}) are stored on the parameter server as a matrix tensor. The topic labels for all the words in one document are stored on the corresponding worker locally in a Python dictionary and are refreshed after each iteration. The topic sampling process loops over each word in a document with \texttt{tf.while\_loop}. Since each document is of variable length, we store the sampled topics in a dynamic-sized \texttt{tf.TensorArray} passed within the \texttt{tf.while\_loop}. The changes in sampled topics are updated to \texttt{ntw} on parameter server via \texttt{tf.scatter\_add}. After each partition of documents is processed, barriers are added on each worker via \texttt{tf.FIFOQueue} for synchronization purpose.

\subsection{Results}

\noindent
\textbf{Efficacy of Cutting Algorithm.}
We begin by examining the utility of the cutting algorithm. Using ten CPU machines, we run FFNN learning (40,000 hidden neurons, batch
size 10,000), W2V learning (100-dimensional embedding) and LDA (1,000 topics), using three different cutting algorithms.  The first
is the simple greedy version of the GQAP solver, as described in Section~\ref{sec:heuristic}.  Second, we use the full solver, but 
rather than taking a probabilistic view of the problem
(Section~\ref{sec:cost-model}), we apply the idea of simply reducing the
number of edges across frames, as these correspond to tables that must be materialized. We call this the ``min-cut'' cutter as
it treats all edges as being equi-weight.
Finally, we evaluate the full algorithm using the cost model of Section~\ref{sec:cost-model}.
We report the per-iteration running time of the various options in Figure \ref{table:cuts}.

To examine the necessity of actually using a frame-based execution, we use ten machines to perform FFNN learning on a relatively
small learning task (10,000 hidden neurons, 
batch size 100).  We unroll 60 iterations of the learning and compare the per-iteration running time 
using the full cutting algorithm along with the cost model of Section~\ref{sec:cost-model}
with a monolithic execution of the entire, unrolled plan.  
The resulting graph has 12,888 relational operators.
The monolithic execution failed during the second iteration.  
The per-iteration running time of the frame-based execution is compared with the running time of the first iteration (under monolithic
execution) in 
Figure \ref{table:full_cut}.

\vspace{5 pt}
\noindent
\textbf{Feed-Forward Networks.}
In the remainder of the experiments, we use the full cutting algorithm with the optimized cost model, along with the frame-based execution.  
On the FFNN learning problem,
we evaluate both the RDBMS and TensorFlow with a variety of cluster sizes (five, ten, and twenty machines) and a wide variety of hidden layer sizes---up to 160,000 neurons.  
{\color{black} Connecting two such layers requires a matrix with 26 billion entries (102 GB).  
Per-iteration execution times are given in Figure \ref{table:ffnn1}.
``Fail'' means that the system crashed.

In addition, we ran a set of experiments where we attempted to achieve the best performance at a \$3-per-hour,
\$7-per-hour, and \$15-per-hour price point using Amazon AWS.
For TensorFlow, at \$3, this was one \texttt{p3.2xlarge} GPU machine and one \texttt{r5.4\-xlarge} CPU machine; at
\$7, it was two \texttt{p3.2xlarge} GPU machines and two \texttt{r5.4xlarge} CPU machines, and
at \$15, it was four \texttt{p3.2xlarge} GPU machines and four \texttt{r5.4xlarge} CPU machines.
SimSQL did about the same using one, two or four \texttt{c5d.18}
\texttt{xlarge} CPU machines (at  \$3, \$7, and \$15,
respectively) as it did using two, five or ten \texttt{g3.4xlarge} GPU machines.
Per-iteration execution times are given in Figure \ref{table:ffnn2}.}

\vspace{5 pt}
\noindent
\textbf{Word2Vec.}
We evaluate both the RDBMS and TensorFlow on a variety of hidden layer sizes, using ten machines.
Per-iteration execution times are given in Figure \ref{table:w2v}.

\vspace{5 pt}
\noindent
\textbf{LDA.}
We next evaluate the RDBMS, TensorFlow, and Spark on LDA, using ten machines and a variety of different model sizes (topic counts).
Sub-step execution times are given in Figure \ref{table:lda}.

\vspace{5 pt}
\noindent
\textbf{Coding Complexity.}
To give the reader an idea of the relative complexity of coding for these systems, in Figure \ref{table:sloc} we give source-line-of-code
counts for each of the various implementations.  Since we implemented all codes from scratch on top of the RDBMS, we had to build
C++ implementations of user-defined functions necessary for the various computations, such as \texttt{crossEntropy\-Derivative}.
We give both SQL and C++ line counts for the RDBMS implementation. 
TensorFlow also has similar C++ code running under the hood.

% new results
\begin{figure}[t]
\begin{center}
\begin{tabular}{|c|c|c|c|}
\hline
Graph Cut Algorithm & FFNN & W2V & LDA\\
\hline
\hline
Fully Optimized Cutter &17:46 & 16:43 & 06:25\\
Min-Cut Cutter &35:29 & 20:53 & 06:21\\
Greedy Cutter &Fail & 25:19 & 06:24\\
\hline
  \end{tabular}
\vspace{10 pt}
\caption{Per iteration running time using frames from various cutting algorithms.}
\vspace{-10 pt}
\label{table:cuts}
\end{center}
\end{figure}
 
\begin{figure}[t]
\begin{center}
\begin{tabular}{|c|c|c|}
\hline
Graph Type & FFNN per-iteration time \\
\hline
\hline
Whole Graph & 05:53:29 \\
Frame-Based & 00:12:53 \\
\hline
  \end{tabular}
\vspace{10 pt}
\caption{Comparing frame-based vs. monolithic (unrolled) plan execution time.}
\vspace{-15 pt}
\label{table:full_cut}
\end{center}
\end{figure} 
 
\begin{figure}[t]
\begin{center}
\begin{tabular}{|c|c|c|}
\hline
\multicolumn{3}{|c|}{FFNN} \\ 
\hline
Hidden Layer Neurons & RDBMS & TensorFlow \\
\hline
\hline
\multicolumn{3}{|c|}{Cluster with 5 workers} \\ 
\hline
10000 &05:39 &01:36\\
20000 &05:46 &03:38\\
40000 &08:30 &09:02\\
80000 &24:52 &Fail\\
160000 &Fail &Fail\\
\hline
\multicolumn{3}{|c|}{Cluster with 10 workers} \\ 
\hline
10000 &04:53 &00:54\\
20000 &05:32 &02:00\\
40000 &07:41 &04:59\\
80000 &17:46 &Fail\\
160000 &44:21 &Fail\\
\hline
\multicolumn{3}{|c|}{Cluster with 20 workers} \\ 
\hline
10000 &04:08 &00:32\\
20000 &05:40 &01:12\\
40000 &06:13 &02:56\\
80000 &12:55 &Fail\\
160000 &25:00 &Fail\\
\hline
  \end{tabular}
\vspace{10 pt}
\caption{Average iteration time for FFNN learning,
using various CPU cluster and hidden layer sizes.}
\vspace{-10 pt}
\label{table:ffnn1}
\end{center}
\end{figure}

\begin{figure}[t]
\begin{center}
\begin{tabular}{|c|c|c|c|}
\hline
\multicolumn{4}{|c|}{FFNN} \\ 
\hline
Hidden Layer & RDBMS & RDBMS & TensorFlow \\
Size & (CPU) & (GPU) & (GPU) \\
\hline
\hline
\multicolumn{4}{|c|}{\$3 per hour budget} \\ 
\hline
10000 &04:50&06:25 &00:24\\
20000 &07:07&07:12&Fail\\
40000 &11:52&11:48 &Fail\\
80000 &16:30&Fail &Fail\\
160000 &Fail & Fail &Fail\\
\hline
\multicolumn{4}{|c|}{\$7 per hour budget} \\ 
\hline
10000 &04:53&04:58 &00:15\\
20000 &05:54&06:08 &Fail\\
40000 &09:32&08:26 &Fail\\
80000 &12:03&17:50 &Fail\\
160000 &Fail & Fail &Fail\\
\hline
\multicolumn{4}{|c|}{\$15 per hour budget} \\ 
\hline
10000 &05:12&5:00 &00:12\\
20000 &05:36 &06:30&Fail\\
40000 &09:08 &08:39&Fail\\
80000 &12:24 &12:20&Fail\\
160000 &39:40 &Fail &Fail\\
\hline
  \end{tabular}
\vspace{15 pt}
\caption{Average iteration time for FFNN learning, maximizing performance at a specific dollar cost.}
\vspace{-10 pt}
\label{table:ffnn2}
\end{center}
\end{figure}

\begin{figure}[t]
\begin{center}
\begin{tabular}{|c|c|c|}
\hline
\multicolumn{3}{|c|}{Word2Vec} \\ 
\hline
Embedding Dimensions & RDBMS & TensorFlow \\
\hline
\hline
100 &00:16:43 (00:01:59) &00:08:03\\
1000 &00:17:05 (00:01:53) &01:14:58\\
10000 &00:29:18 (00:01:53) &Fail\\
\hline
  \end{tabular}
\vspace{10 pt}
\caption{Per iteration running time for Word2Vec.  The time in parens (for the RDBMS) is the time required to execute the code that
generate each batch.
}
\vspace{-15 pt}
\label{table:w2v}
\end{center}
\end{figure}

\begin{figure}[t]
\begin{center}
\begin{tabular}{|c|c|c|c|}
\hline
\multicolumn{4}{|c|}{Collapsed LDA} \\ 
\hline
Number of Topics & RDBMS & TensorFlow & Spark \\
\hline
\hline
1000 &00:06:25 &00:05:06 &00:00:39\\
5000 &00:06:54 &00:25:22 &00:03:03\\
10000 &00:07:05 &00:52:35 &00:06:39\\
50000 &00:08:32 &04:51:51 &00:55:27\\
100000 &00:09:58 &Fail &01:42:35\\
\hline
  \end{tabular}
\vspace{10 pt}
\caption{Average sub-step runtime of collapsed LDA on 10 machines with varied numbers of topics. Format is HH:MM:SS.
}
\vspace{-10 pt}
\label{table:lda}
\end{center}
\end{figure}

\begin{figure}[t]
\begin{center}
\begin{tabular}{|l|c|c|c|}
\hline
& FFNN & Word2Vec & LDA \\
\hline
\hline
RDBMS SQL & 206 & 140 & 135 \\
RDBMS C++ & 324 & 334 & 641 \\
RDBMS total & 530 & 474 & 776 \\
\hline
\hline
TensorFlow Python & 331 & 196 & 227 \\
\hline
\hline
Spark Java & NA & NA & 424 \\
\hline
  \end{tabular}
\vspace{10 pt}  
\caption{Source lines of source code for each of the implementations.}
\vspace{-15 pt}
\label{table:sloc}
\end{center}
\end{figure}

\subsection{Discussion}

\begin{figure}[th!]
\includegraphics[width=8.7cm]{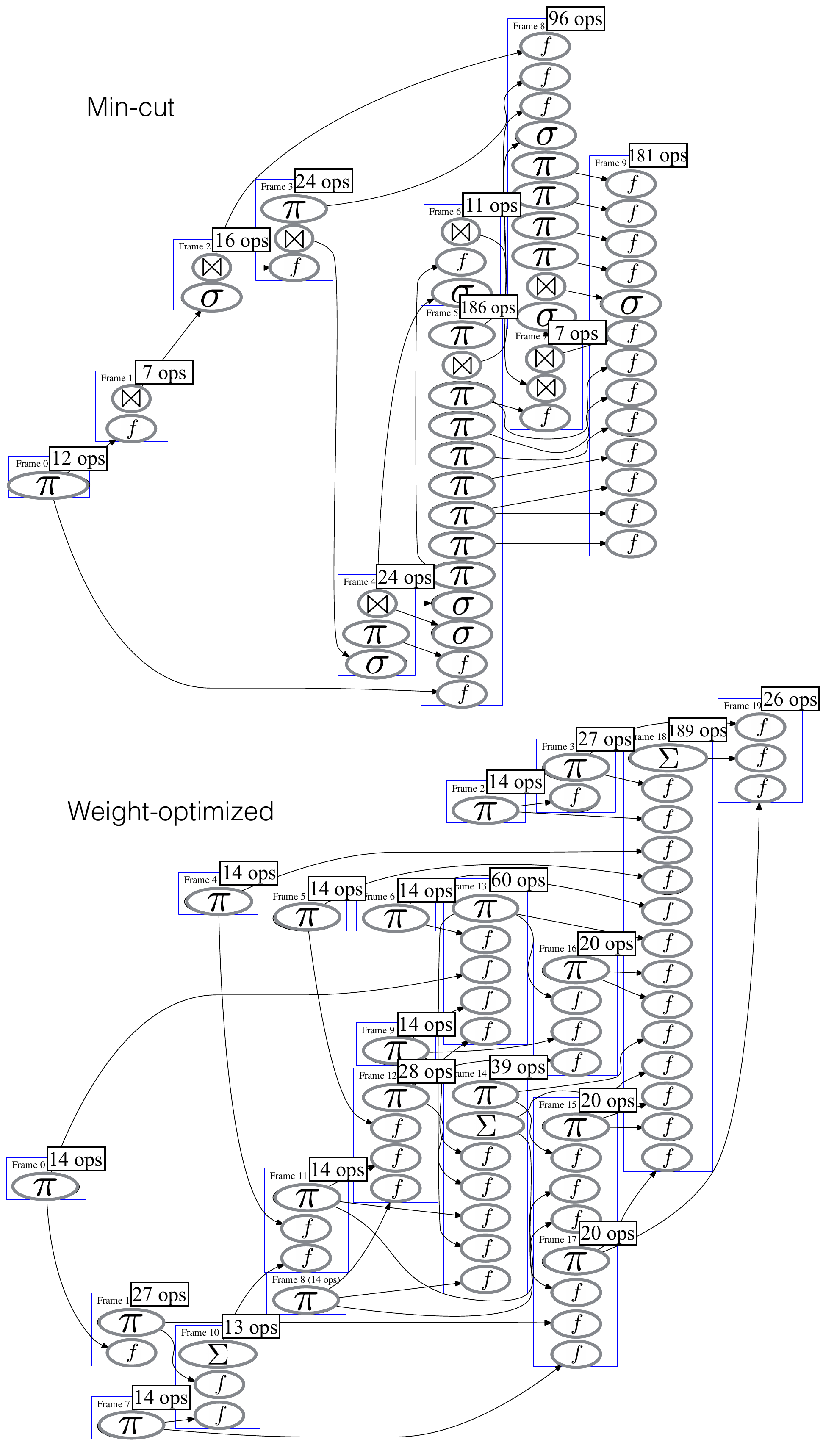}
\vspace{10 pt}
 \caption{Frames created from one iteration of FFNN learning using the min-cut and weight-optimized GQAP formulations.}
 \label{fig:two-graphs}
\end{figure}

{\color{black}
\textbf{Graph cutting.}
SimSQL was
unable to handle the 12,888 operators all together in the FFNN plan, resulting in a running time that was around 28$\times$ longer
than frame-based execution (see Figure \ref{table:full_cut}).  

Figure \ref{table:cuts} shows that, especially for FFNN learning, the full cutting algorithm and cost model is a necessity.
To illustrate how the frames generated from the weight-optimized cutter differ from the min-cut version of the GQAP,
we present Figure \ref{fig:two-graphs}
which shows the set of frames obtained using these two options to cut an unrolling of a single iteration of FFNN learning.  In this graph, we show the relational operators that accept input
into each frame and produce output from each frame.  To represent the relational operations we use 
$\pi$: projection, $\Join$: join, $f$: map, $\Sigma$: aggregate,
$\sigma$: selection.  There are other operator types, but those never produce/process frame IO.  Examining the plots, there
are two obvious differences.  First, the min-cut produces fewer and larger frames, as fewer frames mean fewer edges
to cut.  Second, in almost every case, the weight-optimized cutter chooses to cut across the output from operations that have multiple
consumers.  There are only very few exceptions to this (the projection in Frame 1 whose output is consumed by frame 10, and the aggregation in Frame 18 whose output is consumed by Frame 19).  This is desirable, as explained in Section 6.3, with multiple consumers
of an operation's output, only one can be pipelined, and the rest \emph{must} be materialized.  Hence it is often costless to cut across
such edges.

\vspace{5 pt}
\noindent
\textbf{FFNN Learning.}
On the CPU clusters (Figure \ref{table:ffnn1}), 
the RDBMS was slower than TensorFlow in most cases, but it scaled well, whereas TensorFlow crashed (due to memory
problems) on a problem size of larger than 40,000 hidden neurons.  

Micro-benchmarks
showed that for the 40,000 hidden neuron problem, all of the matrix operations required for an
iteration of FFNN learning took 6 minutes, 17 seconds on a single machine.  
Assuming a perfect speedup, on five machines, learning should take just 1:15 per iteration.
However, the RDBMS took 8:30, and TensorFlow took 9:02.  This shows that both systems incur significant overhead, at least at such a large
model size.
SimSQL, in particular, requires a total
of 61 seconds per FFNN iteration just starting up and tearing down Hadoop jobs.  As the system uses Hadoop, each intermediate result that cannot be pipelined must be written to disk, causing a significant amount of IO.
A faster database could likely lower this overhead significantly.

On a GPU (Figure \ref{table:ffnn2}) 
TensorFlow was very fast, but could not scale past 10,000 neurons.  The problem is that when using a GPU, all data in the computational 
graph must fit on the GPU; TensorFlow is not designed to use CPU RAM as a buffer for GPU memory.  The result is that past 10,000 neurons
(where one weight matrix is 4.8GB), GPU memory is inadequate and the system fails.

Our GPU support in SimSQL did not provide much benefit, for a few reasons.  First, the AWS GPU machines do not
have attached storage, which means that moving to GPU machines leads to all of the disk read/writes incurred by Hadoop happening
over network attached storage (compare with CPU hardware, which had a fast, attached solid-state drive).
Second, as discussed above, SimSQL's overhead beyond pure CPU time for matrix operations is high enough that 
reducing the matrix time further using a GPU was ineffective.

\vspace{5 pt}
\noindent
\textbf{Word2Vec and LDA Learning.}
While FFNN learning plays to TensorFlow's strengths, the system was at a disadvantage for these learning problems compared to an RDBMS.
Both have very large models.  Word2Vec has two matrices of size $10^6 \times d$ for embedding dimensionality $d$; for $d = 10^4$, this is
80GB in size.  To implement negative sampling (which avoids updating all of the weights in this matrix), we need to sample 64 negative words per word pair 
to process a document (consisting of
about 1240 word pairs) .  Each of these 79 thousands of 
samples per document 
generates a separate 
request to the TensorFlow parameter server, where the parameter server extracts a column from a large weight matrix.
These requests are very expensive in TensorFlow, making it slow.  In contrast, the RDBMS implementation simply joins the large
weight matrix (stored as a table of column vectors) with the 79 thousand requested samples, which is fast.

A similar phenomenon happens in LDA learning.  It is necessary to store, on a parameter server, information about how all of the words
in all of the documents are assigned to topics.  This information must be requested during learning.  
Again, these requests are very expensive.  SimSQL handles these requests in bulk, via joins.}

\section{Related Work}
\label{sec:related_work}

\noindent
\textbf{Distributed learning systems.}
The parameter server architecture \cite{Smola:2010, Li:2014:SDM:2685048.2685095} was proposed to provide scalable, 
parallel training for machine learning models. A parameter server consists of two components: 
a parameter server (or key-value store) and a set of workers who repeatedly access and update the model parameters.

DistBelief \cite{Dean:2012} is a framework that targets on training large, deep neural networks on a number of machines. 
It utilizes a parameter-server-like architecture, where the model parallelism is enabled by 
distributing the nodes of a neural network across different machines. 
While the efficacy of this architecture was tested 
on two optimization algorithms (Downpour SGD and Sandblaster L-BFGS), it is unclear
precisely what support DistBelief provides for declarative or automated model parallelism; for example, 
the DistBelief paper did not describe how the matrix-matrix multiplication needed to compute activations is implemented
if the two matrices are partitioned across a set of machines (as \cite{Dean:2012} implied). 

Tensorflow \cite{tensorflow, Abadi:2016} utilizes a similar strategy. 
Although it provides some functions (e.g., \texttt{tf.}\-\texttt{nn.}\-\texttt{embedding}\-\texttt{\_lookup}) that
allow parallel model updates (this function is used in Word2Vec), support for more complex parallel model updates is 
limited. 
For example, TensorFlow does not supply a distributed matrix-matrix multiplication.  
We note that either of these system \emph{could} supply a distributed matrix multiplication as a library call---there is
nothing preventing the use of a tool such as ScaLAPACK \cite{blackford1997scalapack} in either system---but this is a very different approach
than the sort of end-to-end optimizable computations described in this paper.

Project Adam \cite{Chilimbi:2014} applies a similar architecture for use in learning convolutional neural networks. 
The models are partitioned vertically across a set of workers, where fully-connected layers and the convolutional 
layers are separated. 
It is unclear that what support Project Adam supplies for computation over very large sets of weights.

MXNet \cite{Chen:2015} is anther recent system that employs a parameter server to train neural networks. 
The authors of MXNet claim the system supports model parallelism. However, its model parallelism support 
is similar to TensorFlow. 
Complex, model-parallel computations  
require using low-level APIs and manual management of the computations and communications. 

Petuum \cite{Xing:2015} is a framework that provides 
data parallelism and (its authors claim) model parallelism support for large-scale machine learning. 
Follow-on work by Zhang \cite{zhang2015poseidon} considered speeding Pentuum for distributed training, using ideas such
as sending weights as soon as they are updated during backpropagation.
It is unclear, however, how Petuum could handle the large feed-forward network tested in this paper.

There are several other systems providing model parallelism \cite{Krizhevsky2014}. 
AMPNet \cite{Gaunt:2017} adds control flow to the execution graph, and supports dynamic control flow by introducing a well-defined intermediate representation. This framework proves to be efficacy for asynchronous model-parallel training by the experiments.  
Coates et al. \cite{Coates2013DeepLW} built a distributed system on a cluster of GPUs based on the COTS HPC technology. 
This system achieved model parallelism by carefully assigning the partial computations of the whole model to each GPU, and utilized MPI for the communication. 

We have built multi-dimensional-recursion implementation on top of SimSQL \cite{cai2013simulation}, which is a distributed analytics database system. 
The system supports linear algebra computations \cite{luo2017scalable}. We extended SimSQL to enable multi-dimensional recursion, and modified its optimizer to make it feasible for large execution graphs. SAP HANA \cite{Farber:2012} and Hyper \cite{Kemper:2011, Neumann:2015} are two in-memory database systems that support both online transaction processing (OLTP) and online analytical processing (OLAP). Some papers \cite{hana_work, Hyper_work} show that relational database systems can provide effective support for big data analytics and machine learning.

In addition to database systems, many dataflow systems have been developed to support distributed, large-scale data analysis and machine learning,
such as Apache Spark \cite{zaharia2010spark}, Apache SystemML \cite{ghoting2011systemml}, Apache Flink \cite{Carbone2015ApacheFS}, 
and so on.
Both Spark and SystemML provide native libraries for deep learning. 
Moreover, there is a set of deep learning frameworks running on top of Spark, such as Deeplearning4j \cite{deep4j} and BigDL \cite{bigdl}.

\vspace{5 pt}
\noindent
\textbf{Special-purpose deep learning tools}.
Theano \cite{bergstra_theano} is a Python library that facilitates computations on multi-dimensional arrays, 
which provides an easy support for writing deep learning algorithms.
Caffe \cite{Jia:2014} is one of the earliest specialized frameworks for deep learning, 
and mainly focusing on applications to computer vision. 
Caffe2 \cite{caffe2} extends Caffe to provide a better support for large-scale, 
distributed model training, as well as the support for model learning on mobile devices. Torch \cite{Collobert2011Torch7AM} is a computational framework in which users can interact with it with the language Lua. 
Its Python version, PyTorch \cite{pytorch}, applies dynamic computation graphs. 
Similar ideas are adopted by DyNet \cite{Neubig:2017} and Chainer \cite{chainer} as well. 
The Microsoft Cognitive Toolkit (previously known as CNTK) \cite{cntk} is a toolkit that can help people use, or build their own deep learning architectures. 
In addition, higher level APIs are developed on top of those aforementioned frameworks to provide more flexibility for programmers. 
For example, Keras \cite{keras} supports TensorFlow and Theano as its backend, and Gluon \cite{gluon} is run on MXNet. 
Theano does support putting independent computations on different GPUs, 
but it does not provide a complete framework for developing general-purpose model parallel computations.

\section{Conclusions and Future Work}

We have argued that a parallel/distributed RDBMS {\color{black} has promise as a backend} for 
large scale ML computations.  
We have considered unrolling recursive
computations into a monolithic compute plan, which is broken into frames that are optimized and executed independently.
We have expressed the frame partitioning problem as
an instance of the GQAP.  

We have shown that when implemented on top of an RDBMS, these ideas result
in ML computations that are model parallel---that is, able to handle large and complex models that need to be distributed
across machines or compute units.  
{\color{black} We have shown that model parallel, RDBMS-based  ML computations
scale well compared to TensorFlow, and that for 
Word2Vec and LDA, the RDBMS-based computations can be faster than TensorFlow.  The RDBMS was slower than TensorFlow for
GPU-based implementations of neural networks, however.  
Though some of this discrepancy was due to the fact that we implemented 
our ideas on top of a research prototype, high-latency Java/Hadoop system, 
reducing that gap is an attractive target for future work.}

% ensure same length columns on last page (might need two sub-sequent latex runs)
\balance

\vspace{5 pt}
%ACKNOWLEDGMENTS are optional
\section{Acknowledgments}
Thanks to anonymous reviewers for
their insightful feedbacks on earlier versions of this
paper. Work presented in this paper has been supported by the DARPA MUSE program, award
No. FA8750-14-2-0270 and by the NSF under grant
Nos. 1355998 and 1409543.

%\begin{appendix}
%
%\input{appendix}
%
%\end{appendix}

\bibliographystyle{abbrv}
\bibliography{recursion}  % vldb_sample.bib is the name of the Bibliography in this case

%APPENDIX is optional.
% ****************** APPENDIX **************************************
% Example of an appendix; typically would start on a new page
%pagebreak

\end{document}